\documentclass[%
 reprint,
superscriptaddress,
 amsmath,amssymb,
 aps,
]{revtex4-2}

\usepackage{siunitx}
\usepackage{graphicx}
\usepackage{dcolumn}
\usepackage{bm}
\usepackage{hyperref}
\usepackage{algorithm,algorithmic}
\usepackage{mathtools}
\usepackage{physics}

\newcommand{\f}[2][]{\mathcal{F}_{#1}\left[#2\right]}
\newcommand{\finv}[2][]{\mathcal{F}_{#1}^{\dagger} \left[#2\right]}
\renewcommand{\vec}[1]{\mathbf{#1}}
\newcommand{\mvec}[1]{\bm{#1}}
\newcommand{\smatrix}[0]{$\mathcal{S}$-matrix}

\usepackage{color, soul}
\definecolor{linkColor}{rgb}{1,0,0}
\definecolor{citeColor}{rgb}{1,0,0}
\hypersetup{pdfborder={0 0 0}, colorlinks=true,urlcolor=linkColor,citecolor=citeColor}

\emergencystretch=\maxdimen
\begin{document}
\title{Reconstructing the Scattering Matrix from Scanning Electron Diffraction Measurements Alone}

\author{Philipp M Pelz}
\email{philipp.pelz@berkeley.edu}
\affiliation{Department of Materials Science and Engineering, University of California Berkeley, Berkeley, CA 94720}
\affiliation{NCEM, Molecular Foundry, Lawrence Berkeley National Laboratory, Berkeley, CA 94720}

\author{Hamish G Brown}
\affiliation{NCEM, Molecular Foundry, Lawrence Berkeley National Laboratory, Berkeley, CA 94720}

\author{Jim Ciston}
\affiliation{NCEM, Molecular Foundry, Lawrence Berkeley National Laboratory, Berkeley, CA 94720}

\author{Scott D Findlay}
\affiliation{School of Physics and Astronomy, Monash University, Clayton VIC 3800, Australia}


\author{Yaqian Zhang}
\affiliation{Department of Materials Science and Engineering, University of California Berkeley, Berkeley, CA 94720}

\author{Mary Scott}
\affiliation{Department of Materials Science and Engineering, University of California Berkeley, Berkeley, CA 94720}
\affiliation{NCEM, Molecular Foundry, Lawrence Berkeley National Laboratory, Berkeley, CA 94720}

\author{Colin Ophus}
\email{cophus@gmail.com}
\affiliation{NCEM, Molecular Foundry, Lawrence Berkeley National Laboratory, Berkeley, CA 94720}

\date{\today}

\begin{abstract}

Three-dimensional phase contrast imaging of multiply-scattering samples in X-ray and electron microscopy is extremely challenging, due to small numerical apertures, the unavailability of wavefront shaping optics, and the highly nonlinear inversion required from intensity-only measurements. In this work, we present a new algorithm using the scattering matrix formalism to solve the scattering from a non-crystalline medium from scanning diffraction measurements, and recover the illumination aberrations. Our method will enable 3D imaging and materials characterization at high resolution for a wide range of materials. 

\end{abstract}
\maketitle

\section{\label{sec:introduction} Introduction}

Phase contrast imaging is widely used in light \cite{pluta1988advanced, clarke2002microscopy}, x-ray \cite{kirz1995soft, mayo2012line}, and electron microscopy \cite{spence1999future, glaeser2013invited},
due to its high efficiency and resolution. By using coherent radiation to illuminate a sample, we can resolve very small changes in a sample's local index of refraction through the interference of the illumination wave fronts that the accumulated phase shifts produce \cite{zernike1935phase}. However, because we can only directly measure the probability density of a illumination wave function (given by the wave intensity, or amplitude squared), phase contrast imaging is a fundamentally nonlinear measurement process: we must indirectly infer the underlying relative phase shifts induced by the sample \cite{ballentine1970statistical}.

Various approximations can make phase contrast microscopy data easier to interpret. The first is by assuming that the sample is a pure phase object, i.e.\ it does not modulate the illumination wave function amplitude directly, and so any variations in the measured intensity can be directly ascribed to changes in the sample's index of refraction \cite{barer1952ii}. However this assumption does not guarantee uniqueness in all cases, due the possibly of phase wrapping \cite{maretzke2015uniqueness}. An even stronger assumption is the weak phase object approximation (WPOA), where the sample's transmission function is assumed to be a small imaginary perturbation a known carrier wave \cite{born2013principles}. When the WPOA holds, the linear relation implied between specimen potential and measured intensity allow constructive and unambiguous solution. Another commonly used simplification in phase contrast microscopy is the projection approximation (PA), where all scattering is assumed to originate from an infinitesimally thin 2D plane \cite{cohen1984estimates, burvall2011phase}. The various different approximations above hold for a wide range of samples of interest and are therefore very useful in practice \cite{vulovic2014use}.

However, phase contrast imaging of many samples cannot be approximated by any of the above assumptions. Transmission electron microscopy (TEM) in particular often violates these assumptions, due to high scattering cross section of electrons with matter \cite{crewe1974thick}. Instead, these scattering processes can typically only be modeled by a framework that includes multiple scattering \cite{bethe1928theory}. 
The equations describing multiple scattering for a paraxial wave function can be approximately solved with the multislice algorithm \cite{cowley1957scattering}, which has also been used as a model for inverse scattering in many experimental configurations in light, X-ray- and electron microscopy. 
While the inverse multislice model has been successfully applied to image thick, multiply scattering specimens in light microscopy \cite{Chowdhury_Chen_Eckert_Ren_Wu_Repina_Waller_2019, Godden_Suman_Humphry_Rodenburg_Maiden_2014,kamilov2015learning,Li_Maiden_2018}, its use in X-ray \cite{Maiden_Humphry_Rodenburg_2012,suzuki2014high,Shimomura_Suzuki_Hirose_Takahashi_2015,Tsai_Usov_Diaz_Menzel_Guizar-Sicairos_2016,Öztürk_Yan_He_Ge_Dong_Lin_Nazaretski_Robinson_Chu_Huang_2018} and electron microscopy \cite{van2012method,Van_den_Broek_Koch_2013,Gao_Wang_Zhang_Martinez_Nellist_Pan_Kirkland_2017,Schloz_Pekin_Chen_Broek_Muller_Koch_2020} has been limited to proof-of-principle demonstrations with less than 10 slices or weakly scattering samples. 
This is mainly due to the fact that the optical systems in X-ray and electron microscopy have relatively small numerical apertures, such that the information recorded from a single view covers only a small fraction of reciprocal space \cite{Tsai_Marone_Guizar-Sicairos_2019,Jacobsen_2018,Xin_Muller_2009}. 
This problem can be overcome either by enforcing strong prior knowledge about the underlying scattering potential in the form of sparsity constraints or the proper choice of slice separation \cite{Schloz_Pekin_Chen_Broek_Muller_Koch_2020}, or by performing tomographic experiments \cite{Gilles_Nashed_Du_Jacobsen_Wild_2018,Du_Nashed_Kandel_Gürsoy_Jacobsen_2020,ren2020multiple}.

Another framework that incorporates multiple scattering is the scattering matrix (\smatrix{}) formalism \cite{fujimoto1959dynamical, sturkey1962calculation}. 
In electron microscopy, the \smatrix{} formalism has been used to efficiently calculate diffraction results with single crystals~\cite{sturkey1962calculation} and for scanning TEM (STEM) experiments \cite{ophus2017fast, brown2019linear}, and to retrieve projected potentials of strongly scattering samples in a two-step approach. First, the \smatrix{} is retrieved from a series of intensity measurements. Second, the projected structure is retrieved. The proposed experimental methods for retrieval of the \smatrix{} from intensity measurements range from measurements with different crystal thicknesses and sample tilts \cite{spence1998direct}, different sample tilts alone \cite{Allen_Josefsson_Leeb_1998,allen2000inversion,Donatelli_Spence_2020}, wavelength variation \cite{Rez_1999}, large-angle rocking beam diffraction \cite{Wang_Pennington_Koch_2016}, and scanning diffraction with a convergent beam \cite{findlay2005quantitative}. Only the last two of these approaches have been experimentally demonstrated \cite{Wang_Pennington_Koch_2016,brown2018structure}, and only on single-crystal structures.

In the visible light wavelengths, \smatrix{} retrieval and subsequent singular value decomposition allows the identification of transmission eigenchannels \cite{Popoff_Lerosey_Carminati_Fink_Boccara_Gigan_2010} in strongly scattering materials and maximization of energy transport \cite{Kim_Choi_Yoon_Choi_Kim_Park_Choi_2012} through the system. Phase retrieval of the \smatrix{} is performed by real-space phase- \cite{Metzler_Sharma_Nagesh_Baraniuk_Cossairt_Veeraraghavan_2017} or amplitude-modulation \cite{Dremeau_Liutkus_Martina_Katz_Schulke_Krzakala_Gigan_Daudet_2015,Rajaei_Tramel_Gigan_Krzakala_Daudet_2016}, 4-phase interferometry \cite{Popoff_Lerosey_Carminati_Fink_Boccara_Gigan_2010}, or full-field Mach-Zehnder interferometry \cite{Yu_Hillman_Choi_Lee_Feld_Dasari_Park_2013} with input- and output channels in the plane-wave basis. The input and output channels of the \smatrix{} are often represented in real-space, achieved by imaging the output plane with a CCD camera.

Our contribution in this work is three-fold: first, we develop the measurement operator to calculate scanning diffraction intensities of arbitrary samples from a given \smatrix{} and derive its adjoint operator. Second, we formulate a phase retrieval algorithm that retrieves the \smatrix{} of arbitrary samples from a series of scanning diffraction measurements with different modulations of the illumination aperture (e.g. a defocus series). Third, we formulate a relaxation of the phase retrieval algorithm for samples that do not require the full \smatrix{} to be reconstructed. 




\begin{figure*}[ht]
    \includegraphics[width=\textwidth]{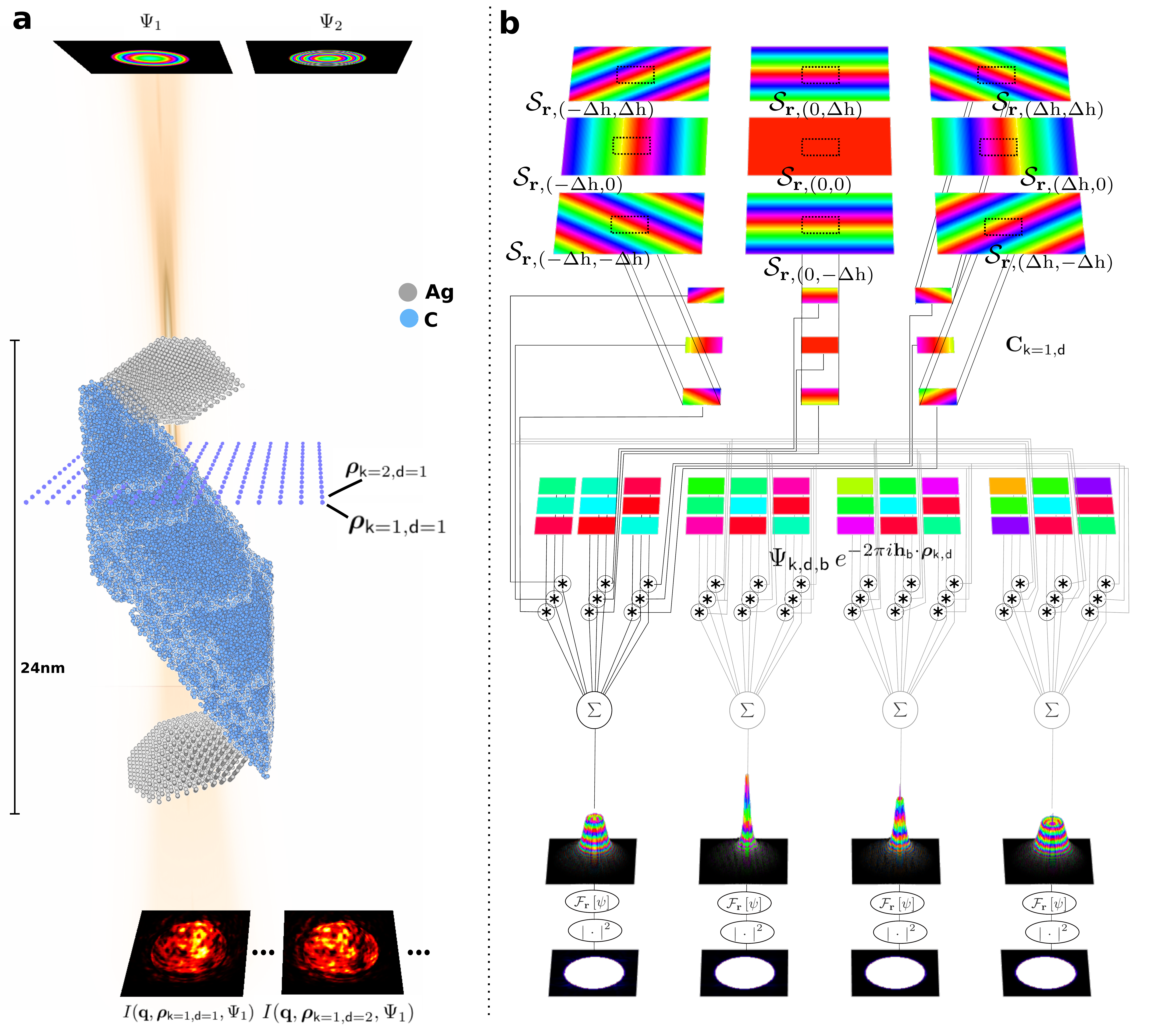}
    \caption{\label{fig:fig1} Measurement scheme for \smatrix{} inversion. (a) A scanning diffraction series of a strongly scattering sample at atomic resolution, where the phases $\Psi_{\mathsf{d}}$ of the probe-forming aperture are varied after each scan, here by changing the defocus. (b) Computational graph of the \smatrix{} measurement operator for $\mathsf{D}=4$ different defocus aberrations. For each scan position, a patch with the size of the diffraction detector ($\mathsf{M}_1 \cross \mathsf{M}_2$) is cropped out of each \smatrix{} beam. Then, each cropped beam is multiplied by the corresponding complex phase factor (indicated by the * operation), depending on the phase and amplitude of beam the illumination aperture $\Psi_{\mathsf{d,b}}$, and the scanning phase $e^{-2 \pi i \vec{h}_{\mathsf{b}}\cdot\bm{\rho}_{\mathsf{k},\mathsf{d}}}$ of the current position. Subsequently, all phase-shifted beams are coherently summed (the $\Sigma$ operator) to form an exit wave. Then the exit wave is propagated to the far-field ($\mathcal{F}_\mathbf{r}$ operation) and measured on the detector.}
\end{figure*}


\section{\label{sec:smatrix_inversion}Reconstructing the \smatrix{}}

\subsection{\label{sec:phase_contrast_background} Theory of phase contrast imaging}

Phase contrast microscopy with coherent light or matter waves defined by the wavefunction $\ket{\psi}_{\vec{r}}$ typically uses a  series of interferometric measurements to invert a partial differential equation of the form
\begin{equation}
    \label{equ:paraxial_schroedinger_equ1}
    i\left[a\,\nabla^2_{\perp} + b\,V(\vec{r^\prime})\right]\ket{\psi}_{\vec{r^\prime}} = \pdv{\ket{\psi}_{\vec{r^\prime}}}{z},
\end{equation}
where $i$ is the imaginary constant, $\nabla^2_{\perp}$ is the two-dimensional Laplace operator, $V(\vec{r^\prime})$ is the three-dimensional potential over the real space coordinates ${\vec{r^\prime} = \left(\vec{r},z\right)}$, and $a$ and $b$ are real-valued constant prefactors. The formal operator solution to this equation for a wave function that has propagated a distance $\Delta z$ through the potential is given by \cite{kirkland2020advanced}, 
\begin{equation}
    \ket{\psi}_{(\vec{r},z+\Delta z)} 
    =
    \exp \left[ 
    ia\,\Delta z \nabla^2_{\perp} + ib\,V_{\Delta z}(\vec{r},z)
    \right]
    \ket{\psi}_{\vec{r^\prime}}.
\end{equation}


In the scattering matrix formalism, the entire process of multiple scattering is modeled by multiplication with the complex-valued linear operator $\mathcal{S}$, 
\begin{equation}
    \ket{\psi}_{out} = \mathcal{S}\ket{\psi}_{in}.
\end{equation}
The \smatrix{} formalism has a wide range of applications in describing the interaction of coherent waves with multiply scattering objects \cite{Rotter_Gigan_2017}. 

All the previously discussed methods for \smatrix{}-retrieval at high resolution have in common that they require a crystalline sample to solve for either the scattering matrix or the structure factors. The interferometric methods developed for light optics rely on the ability to precisely manipulate phases and/or amplitudes of the \smatrix{} input channels and such precise control of the electron and X-ray optics is not yet feasible. In the following section, we describe our iterative reconstruction scheme from scanning diffraction measurements for \smatrix{}-retrieval.

\subsection{\label{sec:multislice_smatrix} A real-space \texorpdfstring{\smatrix{}}{S-matrix} measurement model}
Previous work for retrieving the \smatrix{} from scanning diffraction measurements modeled the formation of the diffraction pattern intensity in the far-field of the sample, given a coherent probe $\ket{\psi}$ at position $\mvec{\rho}$,
\begin{equation}
\label{equ:fourier_shift_theorem}
    \ket{\psi}_{\vec{r}-\mvec{\rho}} = \sum_{|\vec{h}|<h_{\text{max}}}\Psi(\vec{h})e^{2 \pi i\vec{h}\cdot(\vec{r}-\mvec{\rho})},
\end{equation}
with an intensity measurement given by \cite{findlay2005quantitative}
\begin{equation}
    \label{equ:phase_shifting_multislice2}
    I(\vec{q},\mvec{\rho},\Psi) = \left|\sum_{|\vec{h}|<h_{\text{max}}}\mathcal{S}_{\vec{q},\vec{h}}\Psi(\vec{h})e^{-2 \pi i\vec{h}\cdot\mvec{\rho}}\right|^2.
\end{equation}
In this work, we use the approximation that the wave function has a finite support after propagating through the specimen potential. To use this approximation as a constraint in an inversion algorithm, we need to represent the \smatrix{} in real space:
\begin{equation}
    \label{equ:phase_shifting_multislice3}
    I(\vec{q},\mvec{\rho},\Psi) = \left|\f[\vec{r}]{\sum_{|\vec{h}|<h_{\text{max}}}\mathcal{S}_{\vec{r},\vec{h}}\Psi(\vec{h})e^{-2 \pi i\vec{h}\cdot\mvec{\rho}}}\right|^2.
\end{equation}
Here $\mathcal{S}_{\vec{r},\vec{h}}$ is the \smatrix{} that maps Fourier-space input coefficents at wave-vectors $\vec{h}$ (we refer to these as the ``beams'' of the \smatrix{}) to real-space output coefficients at positions $\vec{r}$.

A previous experiment \cite{brown2018structure} used Eq.~\ref{equ:phase_shifting_multislice2} and a series of defocus modulations to retrieve the phases of $\mathcal{S}_{\vec{q},\vec{h}}$ for a set of $\vec{h}$ vectors separately, and then used symmetry relations of $\mathcal{S}_{\vec{q},\vec{h}}$ to find the relative phases between the different \smatrix{} columns. Whereas that approach is only valid for crystalline samples, we use only self-consistency in the measured data and retrieve all amplitudes and phases of $\mathcal{S}_{\vec{r},\vec{h}}$ simultaneously. We also introduce a real-space compactness constraint on the scattered probes produced by the scattering matrix, equivalent to the method of Fourier-interpolating the \smatrix{} \cite{ophus2017fast}.  We introduce the cropping operator,
\begin{equation}
 \mathbf{C}_{\mvec{\rho},\mvec{\Delta}}(\vec{r}) = \begin{cases}
 1 & \mbox{if } r_x - \rho_x \leq \Delta_x/2 \text{ and } r_y - \rho_y \leq \Delta_y/2 \\
 0 & \mbox{otherwise}
 \end{cases}
\end{equation}

a two dimensional rectangular function of width $\mvec{\Delta}$ centered about each probe scan position $\mvec{\rho}$, which transforms Eq.~\ref{equ:phase_shifting_multislice2} into
\begin{equation}
\label{equ:cropped_multislice}
    I(\vec{q},\mvec{\rho},\Psi) = \left|\f[\vec{r}]{\sum_{|\vec{h}|<h_{\text{max}}}\left[\mathbf{C}_{\mvec{\rho},\mvec{\Delta}}(\vec{r})\mathcal{S}_{\vec{r},\vec{h}}\right]\Psi(\vec{h})e^{-2 \pi i\vec{h}\cdot\mvec{\rho}}}\right|^2.
\end{equation}
The fact that the cropping operator acts on all \smatrix{} beams equally leads to a self-consistent solution when measurements are taken with overlapping probe positions. 
\subsection{\label{sec:smatrix_phase_retrieval} Phase retrieval of the \texorpdfstring{\smatrix{}}{S-matrix}}

We now describe an algorithm to retrieve all amplitudes and phases of $\mathcal{S}_{\vec{r},\vec{h}}$ simultaneously, given a set of phase modulations $\{\chi_{\mathsf{d}}(\vec{h})\}_{\mathsf{d}=1,...,\mathsf{D}}$ of the probe-forming aperture, using only self-consistency in the measured data. Let the detector be sampled with $\mathsf{M}_1 \cross \mathsf{M}_2$ pixels. We perform a scan with $\mathsf{K}$ positions and \textsf{D} different probes and label a single position with \textsf{k} and a single defocus with \textsf{d}. Then the measured intensities have the dimension $\mathbf{I} \in \mathbb{R}^{\mathsf{K} \cdot \mathsf{D} \cdot \mathsf{M}_1 \cdot \mathsf{M}_2}$. For ease of notation, we enumerate all \textsf{B} samples in $|\vec{h}| < h_{max}$ with indices $\mathsf{b} = 1, ..., \mathsf{B}$. The \smatrix{} measurement operator maps the \textsf{B} beams of the $\mathcal{S}$-matrix of sampled on a discrete grid of $\mathsf{N}_1 \cross \mathsf{N}_2$ pixels and the $\mathsf{D}$ probes to $\mathsf{K}\cdot\mathsf{D}$ diffraction patterns of size $\mathsf{M}_1 \cross \mathsf{M}_2$. ${\mathcal{A} : \mathbb{C}^{\mathsf{B} \cross \mathsf{N}_1 \cross \mathsf{N}_2} \cross \mathbb{C}^{\mathsf{D} \cross \mathsf{M}_1 \cross \mathsf{M}_2} \rightarrow \mathbb{C}^{\mathsf{K}\mathsf{D}\mathsf{M}_1 \mathsf{M}_2}}$. For better readability, we first define the measurement operator for position $\mathsf{k}$ and probe $\mathsf{d}$: ${\mathcal{A}_{\mathsf{k},\mathsf{d}} : \mathbb{C}^{\mathsf{B} \cross \mathsf{N}_1 \cross \mathsf{N}_2} \cross \mathbb{C}^{\mathsf{M}_1 \cross \mathsf{M}_2} \rightarrow \mathbb{C}^{\mathsf{M}_1 \cdot \mathsf{M}_2}}$:
\begin{equation}
    \mathcal{A}_{\mathsf{k},\mathsf{d}}(\mathcal{S}, \Psi_{\mathsf{d}}) := \left[\f[\vec{r}]{\sum^{\mathsf{B}}_{\mathsf{b = 1}}\Psi_{\mathsf{d,b}}\,e^{-2 \pi i \vec{h}_{\mathsf{b}}\cdot\bm{\rho}_{\mathsf{k},\mathsf{d}}}\left[\mathbf{C}_{\mathsf{k},\mathsf{d}}\mathcal{S}\right]_{\mathsf{b}}}\right]^{V},
\end{equation}
where $[\cdot]^V$ is a vectorization from 2D to 1D. We have also introduced the linear cropping operator ${\mathbf{C}_{\mathsf{k},\mathsf{d}} := \mathbf{C}_{\bm{\rho}_{\mathsf{k},\mathsf{d}}}: \mathbb{C}^{\mathsf{B} \cross \mathsf{N}_1 \cross \mathsf{N}_2} \rightarrow \mathbb{C}^{\mathsf{B} \cross \mathsf{M}_1 \cross \mathsf{M}_2}}$, which extracts a real-space patch of size $\mathsf{M}_1 \cross \mathsf{M}_2$ from each beam of a given \smatrix{} at the position with index \textsf{k} for the phase modulation \textsf{d}. 
The measurement operator for the full experiment is just the operators for each probe and position stacked on top of each other: $\mathcal{A}(\mathcal{S}, \Psi)=\left[\mathcal{A}_{\mathsf{1},\mathsf{1}}(\mathcal{S}, \Psi_{\mathsf{1}}), \mathcal{A}_{\mathsf{2},\mathsf{1}}(\mathcal{S}, \Psi_{\mathsf{1}}), ..., \mathcal{A}_{\mathsf{K},\mathsf{D}}(\mathcal{S}, \Psi_{\mathsf{D}})\right]^T$
We can then write the forward model for the measured intensities of a series of $\mathsf{D}$ scanning diffraction experiments taken with different probes as 
\begin{equation}
\label{equ:forward_model}
\mathbf{y}=\left|\mathcal{A}(\mathcal{S}, \Psi)\right|^2.
\end{equation}
Given this forward model and a set of intensity measurements $\mathbf{I}$ we can formulate the phase retrieval problem for blind $\mathcal{S}$-matrix inversion as 
\begin{eqnarray}
    &&  
    \mathrm{Find}\quad\mathcal{S} \in \mathbb{C}^{\mathsf{B} \cross \mathsf{N}_1 \cross \mathsf{N}_2}\quad\mathrm{and}\quad\Psi \in \mathbb{C}^{\mathsf{D} \cross \mathsf{M}_1 \cross \mathsf{M}_2}\quad \nonumber \\
    && 
    \textrm{Subject to} \quad\left|\mathcal{A}(\mathcal{S}, \Psi)\right|^2 = \mathbf{I} \nonumber.
\end{eqnarray}


If the wave functions $\Psi$ are known, the problem of finding $\mathcal{S}$ from a set of measurements $\mathbf{I}$ is a classical phase retrieval problem. There is a rich history of a algorithmic developments to solve the phase retrieval problem. Historically the first were algorithms based on alternating projections onto non-convex constraint sets \cite{Miao_Sayre_Chapman_1998,Fienup_1982,Shechtman_Eldar_Cohen_Chapman_Miao_Segev_2015}. Since these algorithms lack theoretical convergence guarantees, more recently convex relaxations were developed \cite{PhaseLift_2013,Waldspurger_d’Aspremont_Mallat_2015} which provide a convergence guarantee, but use a prohibitive amount of memory. More recently, Bayesian accelerated gradient methods \cite{Bostan_Soltanolkotabi_Ren_Waller_2018} and methods based on the alternating direction method of multipliers (ADMM) \cite{nikitin2019photon} have become popular. Since the wave functions $\Psi_{\mathsf{d}}$ are usually not known precisely in advance, the problem turns into multi-objective optimization. Additionally, in the presence of noise, it is beneficial to the reconstruction quality to include the noise model of the detector in the optimization. Since most advanced detectors in X-ray and electron microscopy are counting detectors, the noise statistics follow a Poisson distribution: ${\mathbf{I}\sim \mathrm{Poisson}(\mathbf{y})}$. Here we choose an amplitude-based cost function as an approximation to the Poisson likelihood, due to its better convergence behaviour and divergence-free derivative \cite{Yeh_Dong_Zhong_Tian_Chen_Tang_Soltanolkotabi_Waller_2015,Fannjiang_Strohmer_2020}:
\begin{equation}
\label{equ:poisson_likelihood}
    \mathcal{D}(\mathbf{y},\mathbf{I}) := \left|\left|\mathbf{y}-\sqrt{\mathbf{I}}\right|\right|_2,
\end{equation}
where $\|\cdot\|_2$ is the $l_2$ norm and $\mathbf{y}$ are far-field amplitudes of the current model.
We use the ADMM algorithm \cite{Parikh_Boyd_2014} to solve the joint optimization problem of $\mathcal{S}$ and $\Psi$. The augmented Langrangian of the \smatrix{} retrieval problem is
\begin{eqnarray}
\label{equ:loss}
    \mathcal{L}_{\beta}(\mathcal{S}, \Psi, \mathbf{z}, \mathbf{\Lambda}) 
    &=&  
    \mathcal{D}({|\mathbf{z}|}) + \Re{\mathbf{\Lambda}^{\dagger}\left(\mathcal{A}(\mathcal{S}, \Psi)-\mathbf{z}\right)} 
    \nonumber \\
    && + \frac{\beta}{2}\|\mathcal{A}(\mathcal{S},  \Psi)-\mathbf{z}\|^2_2,
\end{eqnarray}
where we have introduced the auxiliary variables ${\mathbf{z} \in \mathbb{C}^{\mathsf{K} \cdot \mathsf{D} \cdot \mathsf{M}_1 \cdot \mathsf{M}_2}}$ and $\mathbf\Lambda \in \mathbb{C}^{\mathsf{K} \cdot \mathsf{D} \cdot \mathsf{M}_1 \cdot \mathsf{M}_2}$, which link the data-loss term with the model-loss term. We seek to solve for $\mathcal{S}$ and $\Psi$ such that $\mathcal{L}(\mathcal{S}, \Psi, \mathbf z,\mathbf\Lambda)$ is minimized:
\begin{equation}
\label{equ:argminS}
    \left(\mathcal{S}^*, \Psi^*, \mathbf z^*, \mathbf\Lambda^*\right) = \operatorname*{arg\,max}_{\Lambda}\operatorname*{arg\,min}_{S, \Psi, \mathbf z}\displaystyle\mathcal{L}(\mathcal{S}, \Psi, \mathbf z, \mathbf\Lambda)
\end{equation}
ADMM decouples the joint problem into subproblems and solves them step by step:
\begin{enumerate}
\item 
    $\Psi^{\textsf{l+1}} = \operatorname*{arg\,min}_{\Psi} \mathcal{L}_{\beta}^{\Psi} := \operatorname*{arg\,min}_{\Psi} \mathcal{L}_{\beta}(\mathcal{S}^{\textsf{l}}, \Psi, z^{\textsf{l}}, \mathbf{\Lambda}^{\textsf{l}})$ 
\item
    $\mathcal{S}^{\textsf{l+1}} = \operatorname*{arg\,min}_{\mathcal{S}} \mathcal{L}_{\beta}^{\mathcal{S}} := \operatorname*{arg\,min}_{\mathcal{S}} \mathcal{L}_{\beta}(\mathcal{S}, \Psi^{\textsf{l+1}}, z^{\textsf{l}}, \mathbf{\Lambda}^{\textsf{l}})$
\item
    $z^{\textsf{l+1}} = \operatorname*{arg\,min}_{z}\mathcal{L}_{\beta}(\mathcal{S}^{\textsf{l+1}}, \Psi^{\textsf{l+1}}, z, \mathbf{\Lambda}^{\textsf{l}})$
\item
     $\mathbf{\Lambda}^{\textsf{l+1}} = \mathbf{\Lambda}^{\textsf{l}} + \beta (z^{\textsf{l+1}} - \mathcal{A}(\mathcal{S}^{\textsf{l+1}},  \Psi^{\textsf{l+1}}))$
\end{enumerate}

\subsection{Subproblems w.r.t. \texorpdfstring{$\Psi$ and $\mathcal{S}$}{Psi and S}}
The subproblems with respect to $\Psi$ and $\mathcal{S}$ both involve the adjoint of the measurement operator $\mathcal{A}$, which for a single measurement is given by
${\mathcal{A}_{\mathsf{k},\mathsf{d}}^{\mathcal{S}\,\dagger} :   \mathbb{C}^{\mathsf{M}_1 \mathsf{M}_2} \rightarrow \mathbb{C}^{\mathsf{B} \cross \mathsf{N}_1 \cross \mathsf{N}_2}}$
\begin{equation}
    \mathcal{A}_{\mathsf{k},\mathsf{d}}^{\mathcal{S}_{\mathsf{b}}\,\dagger}(\mathbf{z}) = \mathbf{C}_{\mathsf{k},\mathsf{d}}^{T}\left[\Psi_{\textsf{d},\textsf{b}}^*e^{2 \pi i\vec{h}_{\mathsf{b}}\cdot\bm{\rho}_{\mathsf{k},\mathsf{d}}}\finv[\vec{q}]{\mathbf{z_{\mathsf{k},\mathsf{d}}}}\right] 
    \label{equ:adjoint_S}
\end{equation}
for a fixed $\Psi$, and ${\mathcal{A}_{\mathsf{k},\mathsf{d}}^{\Psi_{\textsf{d},\textsf{b}}\,\dagger} :  \mathbb{C}^{\mathsf{M}_1 \cdot \mathsf{M}_2} \rightarrow \mathbb{C}^{\mathsf{M}_1 \cross \mathsf{M}_2}}$
\begin{eqnarray}
    \mathcal{A}_{\mathsf{k},\mathsf{d}}^{\Psi_{\textsf{d},\textsf{b}}\,\dagger}(\mathbf{z}) =&& \frac{1}{\mathsf{M_1}\mathsf{M_2}} \sum_{\mathsf{m_1}}^{\mathsf{M_1}}\sum_{\mathsf{m_2}}^{\mathsf{M_2}}\nonumber \\&&\left[\sum_{\mathsf{k=1}}^{\textsf{K}} \left[\mathbf{C}_{\mathsf{k},\mathsf{d}}\mathcal{S}\right]^*_{\mathsf{b}}e^{2 \pi i\vec{h}_{\mathsf{b}}\cdot\bm{\rho}_{\mathsf{k},\mathsf{d}}}\finv[\vec{q}]{\mathbf{z}_{\mathsf{k},\mathsf{d}}}\right]_{\mathsf{m_1},\mathsf{m_2}} 
    \label{equ:adjoint_Psi}
\end{eqnarray}
for a fixed $\mathcal{S}_{\mathsf{b}}$. We solve the subproblems with respect to $\Psi$ and $\mathcal{S}$ with gradient descent. 
\begin{eqnarray}
    \Psi^{\textsf{l+1}} = \Psi^{\textsf{l}} + \gamma_1 \pdv{\mathcal{L}_{\beta}^{\Psi}}{\Psi}\\
    \mathcal{S}^{\textsf{l+1}} = \mathcal{S}^{\textsf{l}} + \gamma_2 \pdv{\mathcal{L}_{\beta}^{\mathcal{S}}}{\mathcal{S}}\,,
\end{eqnarray}
where $\gamma_1,\gamma_2\in \mathbb{R}$ are gradient descent step sizes. We found that one gradient step per iteration is usually enough for fast convergence.
The gradient is given by
\begin{eqnarray}
    \pdv{\mathcal{L}_{\beta}^{\Psi}}{\Psi_{\mathsf{d},\mathsf{b}}} =&& \beta\, \mathcal{A}_{\mathsf{k},\mathsf{d}}^{\Psi_{\textsf{d},\textsf{b}}\,\dagger}(\mathbf{z}^{\textsf{l}}-\mathcal{A}_{\mathsf{k},\mathsf{d}}(\mathcal{S},  \Psi_{\mathsf{d}})-\frac{\mathbf{\Lambda}^{\textsf{l}}}{\beta})\\
    \pdv{\mathcal{L}_{\beta}^{\mathcal{S}}}{\mathcal{S}_{\mathsf{b}}} =&& \beta \sum_{\mathsf{k=1}}^{\textsf{K}}\sum_{\mathsf{d=1}}^{\textsf{D}} \mathcal{A}_{\mathsf{k},\mathsf{d}}^{\mathcal{S}_{\mathsf{b}}\,\dagger}(\mathbf{z}^{\textsf{l}}-\mathcal{A}_{\mathsf{k},\mathsf{d}}(\mathcal{S},  \Psi_{\mathsf{d}})-\frac{\mathbf{\Lambda}^{\textsf{l}}}{\beta}).
\end{eqnarray}

See the Appendix B for a detailed derivation.
\subsection{Subproblem w.r.t. z}
The subproblem w.r.t. \textbf{z} was solved elsewhere \cite{Wen_Yang_Liu_Marchesini_2012}. The solution is
\begin{equation}
    \mathbf{z}^{\textsf{l+1}} = \frac{\mathrm{sgn}(\mathbf{\hat{z}})\left[\sqrt{\mathbf{I}} + \beta |\mathbf{\hat{z}}|\right]}{(1+\beta)}.
\end{equation}
The full ADMM algorithm is then given as:
\begin{algorithm}[H]
	\caption{\small{Joint \smatrix{} and probe retrieval via ADMM}}
	\begin{flushleft}
	\textbf{Input:} \\
	measured intensities $\mathbf{I} \in \mathbb{R}^{\mathsf{K} \cross\mathsf{D} \cross\mathsf{M}_1 \cross\mathsf{M}_2}$\\
	scan positions $\mvec{\rho} \in \mathbb{R}^{\mathsf{K}\cross\mathsf{D}\cross 2}$\\
	initial Fourier space probe phases $\mvec\chi^0 \in \mathbb{C}^{\mathsf{D}\cross\mathsf{B}}$\\
	step sizes $\gamma_1, \gamma_2, \beta \in \mathbb{R}$\\
	\textbf{Initialize:} \\
	set $(\mathsf{N}_1,\mathsf{N}_2)=\lceil\frac{\mathrm{max}(\vec{r}_s)+\mathsf{M}}{\mathsf{M}}\rceil\cdot \mathsf{M}$ such that the plane waves $e^{i\vec{h}\cdot\vec{r}}$ have periodic boundary conditions\\
	calculate $\mathbf{I}^{mean} = \frac{1}{K}\sum_\mathsf{k=1}^{\mathsf{K}}\mathbf{I}_{\mathsf{k}}$ and \\
	$a_{max} = \mathrm{max}\{||\mathbf{I}_{\mathsf{k}}||_1 \forall \mathsf{k}=\{1, ... , K\}\}$\\
	$\Psi^0 \gets \frac{a_{max}}{\sqrt{||\mathbf{I}^{mean}||_1}}\mathbf{I}^{mean}e^{i\mvec\chi^0}$\\
	$\mathcal{S}^0_{\mathsf{b}}\gets e^{i\vec{h}_{\mathsf{b}}\cdot\vec{r}}, \mathcal{S} \in \mathbb{C}^{\mathsf{B}\cross\mathsf{N}_1\cross \mathsf{N}_2}$\\
	$\mathbf\Lambda = \vec{0}, \mathbf z = \vec{0}$
	\end{flushleft}
	\textbf{Run:}
	\begin{algorithmic}[1]
		\FOR{$\mathsf l=0$ to $\mathrm{L}$}
		\STATE $\mathbf{\hat{z}} = \mathbf z^{\textsf{l}} + \frac{\mathbf{\Lambda}^{\textsf{l}}}{\beta}$
		\STATE $ \Psi^{\textsf{l+1}} \gets \Psi^{\textsf{l}} + \gamma_1 \cdot \pdv{\mathcal{L}_{\beta}^{\Psi}}{\Psi}\left(\mathcal{S}^{\textsf{l}}, \Psi^{\textsf{l}}, \mathbf{\hat{z}}\right)$
		\STATE $ \mathcal{S}^{\textsf{l+1}} \gets \mathcal{S}^{\textsf{l}} + \gamma_2 \cdot \pdv{\mathcal{L}_{\beta}^{\mathcal{S}}}{\mathcal{S}}\left(\mathcal{S}^{\textsf{l}}, \Psi^{\textsf{l+1}}, \mathbf{\hat{z}}\right)$
		\STATE $\mathbf{\hat{z}} = \mathcal{A}(\mathcal{S}^{\textsf{l+1}},  \Psi^{\textsf{l+1}}) - \frac{\mathbf\Lambda^{\textsf{l}}}{\beta}$
		\STATE $\mathbf z^{\textsf{l+1}} \gets \frac{\mathrm{sgn}(\mathbf{\hat{z}})\left[\sqrt{\mathbf{I}} + \beta |\mathbf{\hat{z}}|\right]}{(1+\beta)}$\\
		\STATE $\mathbf\Lambda^{\textsf{l+1}} \gets \mathbf\Lambda^{\textsf{l}} + \beta (\mathbf z^{\textsf{l+1}} - \mathcal{A}(\mathcal{S}^{\textsf{l+1}},  \Psi^{\textsf{l+1}}))$ 
		\ENDFOR
	\end{algorithmic}
	\textbf{Output:}
	$\mathcal{S}^* = \mathcal{S}^{\mathrm{L}}$
	\label{alg:smatrix}
\end{algorithm}


\section{Simulated \texorpdfstring{\smatrix{}}{S-matrix} Phase Retrieval}

In this section, we use forward simulations to validate our \smatrix{} phase retrieval algorithm. We also examine the algorithm dependence on the sampling density and calibration.

\subsection{\label{sec:sim} Sampling and calibration dependence}

\begin{figure*}[ht!]
\includegraphics[width=\textwidth]{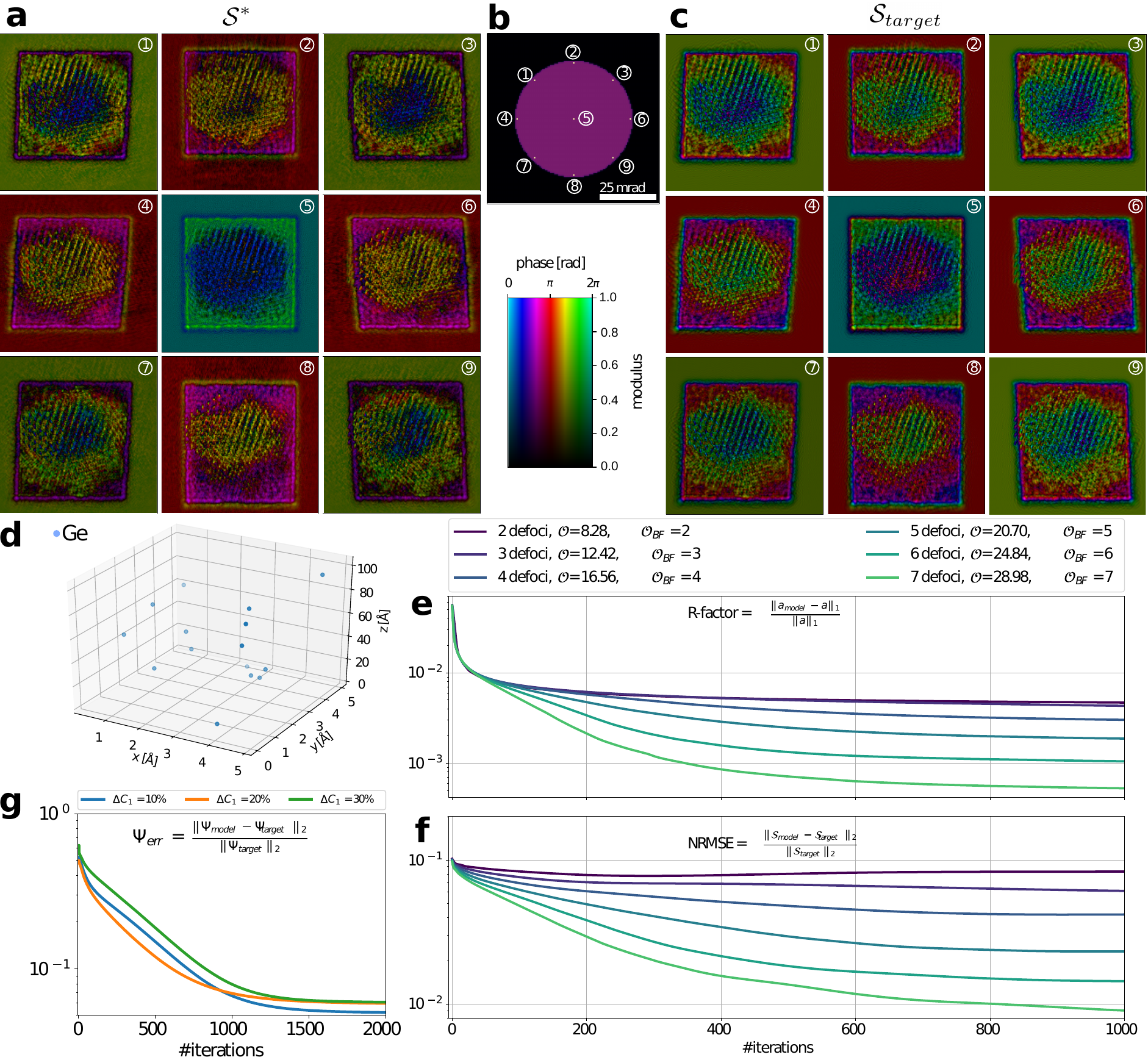}
\caption{\label{fig:fig2} (a) Simulated experiment with (b) the probe-forming aperture used for simulating the experiment shown in Fig. \ref{fig:fig1} (a). Selected beams numbered in (b) are shown from the reconstructed \smatrix{} in (a) and the ground-truth \smatrix{} in (c). The skew effect of the exit waves in different beams comes from the three-dimensional structure, and is a parallax effect of the different propagation directions of the beams. (d) Test sample of randomly distributed Germanium atoms. (e) R-factor vs number of iterations for different numbers of defoci and oversampling rates used in the simulations. (f) Normalized root mean square error of the model \smatrix{} vs number of iterations for different numbers of defoci and oversampling rates. h) Mean probe error vs number of iterations for defocus miscalibration levels of \SI{10}{\percent}, \SI{20}{\percent} and \SI{30}{\percent} of the defocus step and random higher order aberrations.}
\end{figure*}

To demonstrate that our algorithm can reconstruct $\mathcal{S}$-matrices of realistic samples, we simulate a 4D-STEM focal series of the sample shown in Fig. \ref{fig:fig1} a), as it may appear in a tomography experiment. The sample contains two decahedral Ag nanoparticles of \SI{3.3}{\nano\meter} diameter, placed on the top and bottom sides of an amorphous carbon substrate, tilted by \SI{67}{}$^{\circ}$, giving it an axial extent of \SI{24}{\nano\meter}. The probe convergence angle is chosen as \SI{26}{\milli\radian} and the electron energy as \SI{300}{\kilo\volt}, resulting in a depth of focus (DOF) of \SI{5.8}{\nano\meter} and a sample depth of \num{4.1}$\times$ DOF. The detector was set to record diffraction signal up to \SI{40}{\milli\radian}, resulting in a sampling grid with steps of \SI{25}{\pico\meter}. The field of view was scanned with \num{129x129} positions on a 2D grid with the half-period resolution. The reconstruction shown in Fig. \ref{fig:fig2} used \num{6} defoci with a step of \SI{4.6}{\nano\meter}, with the first defocus at the top of the sample. The detector size was set to \num{128x128} pixels, yielding an angular resolution of \SI{0.31}{\micro\radian} and \smatrix{} dimensions of ${\mathcal{S} \in \mathbb{C}^{\mathsf{5973} \cross \mathsf{256} \cross \mathsf{256}}}$. 

We ran Algorithm \ref{alg:smatrix} for 500 iterations, utilizing 48 NVIDIA V-100 GPUs. After 200 minutes, the reconstruction converged to an normalized root mean square error (NRMSE) of \SI{4}{\percent} and an R-factor of \SI{0.1}{\percent}. Nine selected \smatrix{} beams from the reconstruction are shown in Fig. \ref{fig:fig2} a, and the ground-truth \smatrix{} is shown in \ref{fig:fig2}c. To investigate the convergence properties under varying number of measurements and calibration errors, we used a smaller test sample, consisting of 16 randomly distributed Germanium atoms in a volume of \SI{5x5x100}{\angstrom}, shown in Fig. \ref{fig:fig2}d. The convergence angle for the following tests was chosen as \SI{30}{\milli\radian}, with a detector spanning \SI{60}{\milli\radian}, and the diffraction patterns were sampled on a \num{20x20} pixel detector, yielding \smatrix{} dimensions of $\mathcal{S} \in \mathbb{C}^{\mathsf{177} \cross \mathsf{60} \cross \mathsf{60}}$, and the defocus step was chosen as \SI{2}{\nano\meter}.

For the following investigations we fix the scan step to Nyquist sampling. First we investigate the converge behaviour with respect to the number of measured defoci. Fig.~\ref{fig:fig2}e and f show the R-factor, and the NRMSE as a function of iterations and number of defoci measured. We define the oversampling factor as
\begin{equation}
    \mathcal{O} = \frac{\text{\# nonzero measurements}}{\text{\# variables in \smatrix{}}},
\end{equation}
and the bright-field oversampling factor as 
\begin{equation}
    \mathcal{O}_{BF} = \frac{\text{\# nonzero measurements in bright-field}}{\text{\# variables in \smatrix{}}}.
\end{equation}
One can see that for 2 defocus measurements, the NRMSE diverges slowly, and for 3 measurements the NRMSE does not converge monotonously with the R-factor. While the oversampling factor $\mathcal{O}$ lies above the number 4 typically needed for successful phase retrieval, the number of phase modulations that each beam receives, $\mathcal{O}_{BF}$, is below the threshold. For this case, a more heterogeneous sample than the crystalline objects considered in previous work, the reconstruction does not stably converge in these cases. This could be due to the small defocus steps used and will be investigated in the future.

We also investigate the dependence of the probe refinement on the level of defocus miscalibration and residual uncorrected probe aberrations. Fig.~\ref{fig:fig2} h) shows the mean errors of 30 reconstructions performed with defocus errors $\Delta C_1$ drawn form a normal distribution with a standard deviation of \SI{10}{\percent}, \SI{20}{\percent} and \SI{30}{\percent} of the defocus step, axial coma with a standard deviation of \SI{100}{\nano\meter}, three-fold astigmatism with a standard deviation of \SI{20}{\nano\meter}, spherical aberration with a standard deviation of \SI{4}{\micro\meter}, and star aberration with a standard deviation of \SI{4}{\micro\meter}. Although convergence takes roughly twice as many iterations \smatrix{}-reconstruction with mis-calibrated aberrations, for all miscalibration values a probe reconstruction error of less than \SI{10}{\percent} was achieved.

\subsection{\label{sec:collapsed_smatrix_phase_retrieval} Reconstructing the projected \texorpdfstring{\smatrix{}}{S-matrix}}

Consider the scattering matrix for a phase object, which is a valid approximation for a thin and weakly scattering sample~\cite{vulovic2014use}, with specimen potential $V(\vec{r})$. The analytic expression for each component will be,
\begin{equation} \label{eq:WPOASmatrix}
    \mathcal{S}_{\vec{r},\vec{h}} = e^{i\sigma V(\vec{r})-2\pi i \vec{h}\cdot\vec{r}}\,.
\end{equation}

So every \smatrix{} component will be the same except for the multiplicative phase ramp of $e^{-2\pi i \vec{h}\cdot \vec{r}}$. 
As we consider thicker, more strongly scattering objects we would expect each component of the \smatrix{} to be increasingly different and we consider the similarity or lack thereof of each of the \smatrix{} components to be an indication of the degree of strong multiple scattering of a sample.
When reconstructing the \smatrix{} from a 4D-STEM dataset the automatic choice for choosing the sampling of beams, the set of $\vec{h}$ vectors, is to match it to the number of pixels within the bright-field disk or aperture function of the STEM probe in the diffraction patterns.
For a fine diffraction space sampling of an object that does not exhibit much multiple scattering this sampling of beams might be highly redundant and we might improve our reconstruction by forcing a more sparse sampling of beams and increasing the ratio of experimental measurements to unknown parameters in our reconstruction.
On the other hand, very thick and strongly scattering samples might require very high sampling of the diffraction patterns for an accurate reconstruction of the \smatrix{}.
While the latter case can only be solved with better sampling in the diffraction plane, for the former case in this section we outline a strategy for choosing a sparser sampling of the input beams $\vec{h}$ that involves partitioning of the bright-field disk into separate ``tiles''.

\begin{figure}[ht!]
\includegraphics[height=0.72\textheight]{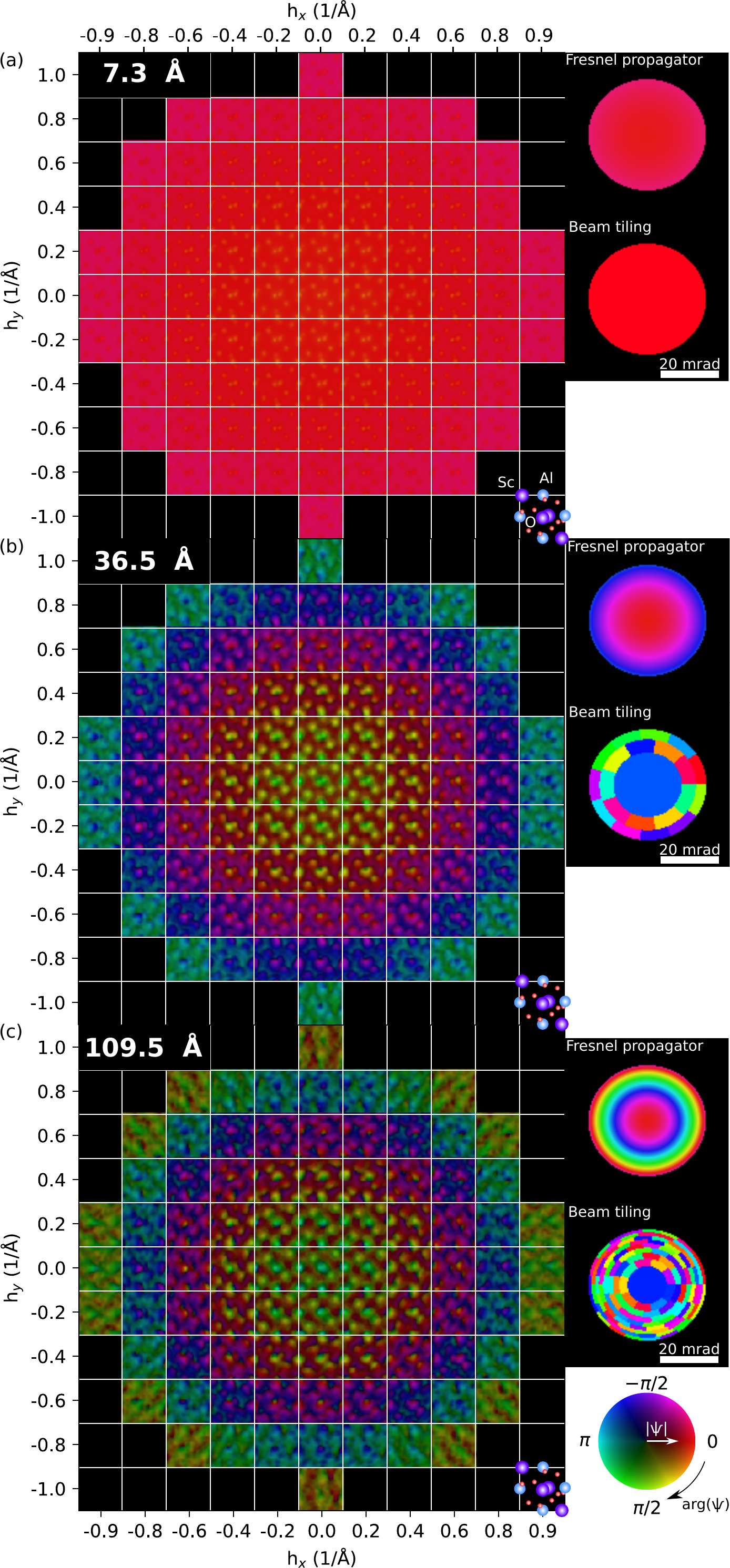}
\caption{\label{fig:fig2b} 
Partitioning of the \smatrix{} beams into separate tiles according to the expected degree of departure from the phase object approximation. In each subplot a subset of the complex components of the scattering matrix (color hue is phase and color saturation is amplitude according to inset colorwheel) are shown for a ScAlO$_3$ crystal with (a) 7.3 \AA, (b) 36.5 \AA\ and (c) 109.5 \AA\ thickness. With increasing thickness there is increasing variability between the different components. The beam partitioning suggested by a phase variance of the Fresnel freespace propagator of $\pi/4$ described in the text is inset on the top right for each of the thicknesses in (a)-(c) where each colour in the bright-field disk corresponds to a seperate $\mathsf{B}_{\text{tile}}$
}
\end{figure}

Shown in Fig.~\ref{fig:fig2b} are the complex values of a subset of \smatrix{} components for a) 7.3 \AA, b) 36.5 \AA\ and c) 109.5 \AA\ thicknesses of an ScAlO$_3$ crystal. 
All beams have been multiplied by the conjugate of the phase ramp that appears in Eq.~\ref{eq:WPOASmatrix}, $e^{2\pi i \vec{h}\cdot{r}}$. 
For Fig.~\ref{fig:fig2b}(a) the difference between beams is minimal so a phase object approximation would be appropriate for this thickness. For Fig.~\ref{fig:fig2b} b)-c) we see increasing variation between beams as the object becomes thicker. 
A partitioning system aims to group these \smatrix{} components by similarity and visual comparison of the \smatrix{} montages with the Fresnel propagator, $\mathcal{P}(h)=\exp(-i\lambda\pi h^2 t)$, for free-space of equivalent thickness of the crystal (shown to the right of each subfigure) suggests that a criterion based on phase variation of a Fresnel free-space propagator might be an effective way of doing this.
We partition the bright-field disk into annular regions $\sqrt{(i-1)\Delta \phi/\lambda\pi t}< h_{i}<\sqrt{i\Delta \phi/\lambda\pi t}$ where ${i\in \mathbb{N}}$ and $\Delta \phi$, the Fresnel propagator phase variance, is a predetermined criterion (we use $\Delta \phi=\pi/4$ in this work).
These regions are further divided azimuthally, with an arclength equal to the radius of the inner-most partition, $\sqrt{\Delta \phi/\lambda\pi t}$.
We represent these partitions with the map $\tau : \{0, ... , \mathsf{B}\} \rightarrow \{0, ... , \mathsf{B_{tile}\}}$ from beams to beam tiles. 
Partitioning according to his criterion is shown to the right where each different color indicates a separate partition of the bright-field disk for each of the thicknesses in Fig.~\ref{fig:fig2b}(a)-(c).
We note finally that this is an approximate criterion only since thickness of an uncharacterised object can only be guessed at based on the intuition of the microscopist and the Fresnel criterion does not take into account the scattering strength per unit volume of the object.
For example we might expect samples containing a high density of heavy (large $Z$) elements to exhibit greater beam to beam variation of the  \smatrix{} components than materials with only small $Z$ numbers.


Partitioning reduces the number of needed measurements by a factor $\mathsf{B_{tile}}/\mathsf{B}$. 
The reconstructed \smatrix{} then has the reduced dimension $\mathcal{S} \in \mathbb{C}^{\mathsf{B_{tile}} \cross \mathsf{N}_1 \cross \mathsf{N}_2}$, and the beam-dependent beam-tilt $e^{-2\pi i\vec{h}\cdot\mvec{r}}$ has to be separated from the \smatrix{} variable to allow the beam-averaging. 
We define therefore the un-tilted \smatrix{} ${\mathcal{S}^{t}_{\vec{r},\vec{h}} := \mathcal{S}_{\vec{r},\vec{h}}e^{-2 \pi i\vec{h}\cdot\vec{r}}}$, which is the new latent-variable in the projected \smatrix{} problem. 
The forward operator becomes
\begin{eqnarray}
    &&
    \mathcal{A}^{t}_{\mathsf{k},\mathsf{d}}(\mathcal{S}^{t}, \Psi_{\mathsf{d}}) :=
    \nonumber \\
    &&
    \left[\f[\vec{r}]{\sum^{\mathsf{B}}_{\mathsf{b = 1}}\Psi_{\mathsf{k,d,b}}\,e^{-2 \pi i \vec{h}_{\mathsf{b}}\cdot\bm{\rho}_{\mathsf{k},\mathsf{d}}}\left[\mathbf{C}_{\mathsf{k},\mathsf{d}}\mathcal{S}^{t}\right]_{\mathsf{\tau(b)}}e^{2\pi i\vec{h}\cdot\mvec{r}}}\right]^{V},
    \nonumber
\end{eqnarray}
and the gradient with respect to $\mathcal{S}^{t}$ is
\begin{eqnarray}
    \pdv{\mathcal{L}_{\beta}^{\mathcal{S}^{t}}}{\mathcal{S}^{t}_{\mathsf{b}}}
    = \frac{\beta}{|Q_{\mathsf{b}}|} \sum_{\mathsf{q} \in Q_{\mathsf{b}}}&&\sum_{\mathsf{k=1}}^{\textsf{K}}\sum_{\mathsf{d=1}}^{\textsf{D}} \nonumber\\
    &&
    \mathcal{A}_{\mathsf{k},\mathsf{d}}^{\mathcal{S}_{\mathsf{b}}\,\dagger}(\mathbf{z}^{\textsf{l}}-\mathcal{A}^{t}_{\mathsf{k},\mathsf{d}}(\mathcal{S}^{t},  \Psi_{\mathsf{d}})-\frac{\mathbf{\Lambda}^{\textsf{l}}}{\beta})
\end{eqnarray}
where we have introduced the set ${Q_{\mathsf{b}} = \{n\,|\,\tau(n)=b \,\forall\, n\,\in \{1, ... , B\} \}}$ of beams that belong to tile $\mathsf{b}$ and $|Q_{\mathsf{b}}|$ is the cardinality of $Q_{\mathsf{b}}$.

\section{Conclusion and Outlook}

We have introduced a new method for \smatrix{} retrieval, that converges without any regularization for samples which span 4 depths of focus and more, and numerical apertures which are experimentally accessible, and can recover aberration miscalibrations of up to \SI{30}{\percent}. We have also introduced a simplified model, projected \smatrix{} inversion, for the case when the sample is thin enough that not every beam that is measured on the detector has to be included in the model. 
In future work, we will compare projected \smatrix{} inversion to mixed-state ptychography and multi-slice ptychography, since both offer alternative methods for moving beyond the simple model of single-mode ptychography. 


The \smatrix{}-retrieval methods developed here could be used for a number of advancements in imaging through and with strongly scattering materials in X-ray and electron microscopy. In combination with adaptive electron optics \cite{Verbeeck_Beche_Muller_Caspary_Guzzinati_Luong_Den_Hertog_2018}, selective focusing through crystalline materials may become possible in a similar vein to light optical experiments \cite{Kong_Silverman_Liu_Chitnis_Lee_Chen_2011}.\\
The retrieved \smatrix{} can be used for depth-sectioning which is robust against multiple scattering. \smatrix{}-retrieval may also form the basis of inverse multi-slice algorithms for phase-contrast tomography in scanning diffraction microscopes. The angular decomposition in the \smatrix{} may be useful for ab-initio angular and transverse alignment of different tilt angles for phase-contrast tomography. This approach may be experimentally more feasible than end-to-end tomographic reconstruction algorithms. Finally, one could think about characterizing amorphous materials from their \smatrix{}. \\
To allow optimal image quality, future refinements of the algorithm could include experimental uncertainties like position errors, and modeling of nuisance parameters like spatial and temporal incoherence, similar to their treatment in ptychographic reconstruction algorithms \cite{Maiden_Humphry_Sarahan_Kraus_Rodenburg_2012,Odstril_Menzel_Guizar_Sicairos_2018,Rana_Zhang_Pham_Yuan_Lo_Jiang_Osher_Miao_2020,Thibault_Menzel_2013,Chen_Odstrcil_Jiang_Han_Chiu_Li_Muller_2020}.

\section*{Acknowledgments}

We thank Tia Pelz, Nicole Morello, and Peter Hosemann for support during the COVID-19 pandemic. Without them, this research would not have been possible. PMP acknowledges financial support from STROBE. HGB and JC acknowledge support from the Presidential Early Career Award for Scientists and Engineers (PECASE) through the U.S. Department of Energy. CO acknowledges support from the Department of Energy Early Career Research Award program. Work at the Molecular Foundry was supported by the Office of Science, Office of Basic Energy Sciences, of the U.S. Department of Energy under Contract No. DE-AC02-05CH11231. This research was partly supported under the Discovery Projects funding scheme of the Australian Research Council (Project No. FT190100619).

\appendix
\section{\label{sec:complexity} Complexity analysis}

Both the forward calculation and the backward calculation have the following computational complexity: per diffraction pattern the forward pass has a complexity of ${O(\mathsf{M_1}^2\mathsf{M_2}^2\mathsf{B}\log(\mathsf{M_1}\mathsf{M_2}))}$ where the factor ${O(\mathsf{M_1}\mathsf{M_2}\log(\mathsf{M_1}\mathsf{M_2}))}$ comes from the fast Fourier transform operation. The forward and backward calculation on the full dataset then have a complexity of ${O(\mathsf{K}\mathsf{D}\mathsf{M_1}^2\mathsf{M_2}^2\mathsf{B}\log(\mathsf{M_1}\mathsf{M_2}))}$. Since the number of beams scales quadratically with the size of the detector, the overall complexity scales with  ${O(\mathsf{K}\mathsf{D}\mathsf{M}^4\log(\mathsf{M}^2))}$ for a square detector of size $\mathsf{M}$. While this might seem intractable for currently available large detectors, it is offset by the fact that $\mathsf{K}\mathsf{D}\mathsf{M}^2$ of these computations are embarrassingly parallel batched complex matrix multiplications and can be carried out very efficiently on commonly available hardware accelerators.

\bibliography{apssamp}

\begin{thebibliography}{74}
\providecommand{\natexlab}[1]{#1}
\providecommand{\url}[1]{\texttt{#1}}
\expandafter\ifx\csname urlstyle\endcsname\relax
  \providecommand{\doi}[1]{doi: #1}\else
  \providecommand{\doi}{doi: \begingroup \urlstyle{rm}\Url}\fi

\bibitem[Pluta(1988)]{pluta1988advanced}
M.~Pluta.
\newblock \emph{Advanced light microscopy}, volume~1.
\newblock Elsevier Amsterdam, 1988.

\bibitem[Clarke and Eberhardt(2002)]{clarke2002microscopy}
A.~Clarke and C.~N. Eberhardt.
\newblock \emph{Microscopy techniques for materials science}.
\newblock Woodhead Publishing, 2002.

\bibitem[Kirz et~al.(1995)Kirz, Jacobsen, and Howells]{kirz1995soft}
J.~Kirz, C.~Jacobsen, and M.~Howells.
\newblock Soft {X}-ray microscopes and their biological applications.
\newblock \emph{Quarterly reviews of biophysics}, 28\penalty0 (1):\penalty0
  33--130, 1995.

\bibitem[Mayo et~al.(2012)Mayo, Stevenson, and Wilkins]{mayo2012line}
S.~C. Mayo, A.~W. Stevenson, and S.~W. Wilkins.
\newblock In-line phase-contrast {X}-ray imaging and tomography for materials
  science.
\newblock \emph{Materials}, 5\penalty0 (5):\penalty0 937--965, 2012.

\bibitem[Spence(1999)]{spence1999future}
J.C.H. Spence.
\newblock The future of atomic resolution electron microscopy for materials
  science.
\newblock \emph{Materials Science and Engineering: R: Reports}, 26\penalty0
  (1-2):\penalty0 1--49, 1999.

\bibitem[Glaeser(2013)]{glaeser2013invited}
R.~M. Glaeser.
\newblock Invited review article: Methods for imaging weak-phase objects in
  electron microscopy.
\newblock \emph{Review of Scientific Instruments}, 84\penalty0 (11):\penalty0
  312, 2013.

\bibitem[Zernike(1935)]{zernike1935phase}
F~Zernike.
\newblock Phase contrast.
\newblock \emph{Z. Tech. Physik.}, 16:\penalty0 454, 1935.

\bibitem[Ballentine(1970)]{ballentine1970statistical}
L.~E. Ballentine.
\newblock The statistical interpretation of quantum mechanics.
\newblock \emph{Reviews of Modern Physics}, 42\penalty0 (4):\penalty0 358,
  1970.

\bibitem[Barer(1952)]{barer1952ii}
R~Barer.
\newblock A vector theory of phase contrast and interference contrast. i.
  positive phase contrast.
\newblock \emph{Journal of the Royal Microscopical Society}, 72\penalty0
  (1):\penalty0 10--30, 1952.

\bibitem[Maretzke(2015)]{maretzke2015uniqueness}
S.~Maretzke.
\newblock A uniqueness result for propagation-based phase contrast imaging from
  a single measurement.
\newblock \emph{Inverse Problems}, 31\penalty0 (6):\penalty0 065003, 2015.

\bibitem[Born and Wolf(2013)]{born2013principles}
M.~Born and E.~Wolf.
\newblock \emph{Principles of optics: electromagnetic theory of propagation,
  interference and diffraction of light}.
\newblock Elsevier, 2013.

\bibitem[Cohen et~al.(1984)Cohen, Schmid, and Chiu]{cohen1984estimates}
H.~A. Cohen, M.~F. Schmid, and W.~Chiu.
\newblock Estimates of validity of projection approximation for
  three-dimensional reconstructions at high resolution.
\newblock \emph{Ultramicroscopy}, 14\penalty0 (3):\penalty0 219--226, 1984.

\bibitem[Burvall et~al.(2011)Burvall, Lundstr{\"o}m, Takman, Larsson, and
  Hertz]{burvall2011phase}
A.~Burvall, U.~Lundstr{\"o}m, P.A.C. Takman, D.~H. Larsson, and H.~M. Hertz.
\newblock Phase retrieval in {X}-ray phase-contrast imaging suitable for
  tomography.
\newblock \emph{Optics express}, 19\penalty0 (11):\penalty0 10359--10376, 2011.

\bibitem[Vulovi{\'c} et~al.(2014)Vulovi{\'c}, Voortman, van Vliet, and
  Rieger]{vulovic2014use}
M.~Vulovi{\'c}, L.~M. Voortman, L.~J. van Vliet, and B.~Rieger.
\newblock When to use the projection assumption and the weak-phase object
  approximation in phase contrast cryo-{EM}.
\newblock \emph{Ultramicroscopy}, 136:\penalty0 61--66, 2014.

\bibitem[Crewe and Groves(1974)]{crewe1974thick}
A.~V. Crewe and T.~Groves.
\newblock Thick specimens in the cem and stem. i. contrast.
\newblock \emph{Journal of Applied Physics}, 45\penalty0 (8):\penalty0
  3662--3672, 1974.

\bibitem[Bethe(1928)]{bethe1928theory}
H.~Bethe.
\newblock The theory of electron diffraction on crystals.
\newblock \emph{Annals of Physics}, 392\penalty0 (17):\penalty0 55--129, 1928.

\bibitem[Cowley and Moodie(1957)]{cowley1957scattering}
J.~M. Cowley and A.~F. Moodie.
\newblock The scattering of electrons by atoms and crystals. {I}. a new
  theoretical approach.
\newblock \emph{Acta Crystallographica}, 10\penalty0 (10):\penalty0 609--619,
  1957.

\bibitem[Chowdhury et~al.(2019)Chowdhury, Chen, Eckert, Ren, Wu, Repina, and
  Waller]{Chowdhury_Chen_Eckert_Ren_Wu_Repina_Waller_2019}
S.~Chowdhury, M.~Chen, R.~Eckert, D.~Ren, F.~Wu, N.~Repina, and L.~Waller.
\newblock High-resolution 3d refractive index microscopy of multiple-scattering
  samples from intensity images.
\newblock \emph{Optica}, 6\penalty0 (9):\penalty0 1211–1219, Sep 2019.
\newblock ISSN 2334-2536.
\newblock \doi{10/ggzqdc}.

\bibitem[Godden et~al.(2014)Godden, Suman, Humphry, Rodenburg, and
  Maiden]{Godden_Suman_Humphry_Rodenburg_Maiden_2014}
T.~M. Godden, R.~Suman, M.~J. Humphry, J.~M. Rodenburg, and A.~M. Maiden.
\newblock Ptychographic microscope for three-dimensional imaging.
\newblock \emph{Optics Express}, 22\penalty0 (10):\penalty0 12513–12523, May
  2014.
\newblock ISSN 1094-4087.
\newblock \doi{10/ggzqfn}.

\bibitem[Kamilov et~al.(2015)Kamilov, Papadopoulos, Shoreh, Goy, Vonesch,
  Unser, and Psaltis]{kamilov2015learning}
U.~S. Kamilov, I.~N. Papadopoulos, M.~H. Shoreh, A.~Goy, C.~Vonesch, M.~Unser,
  and D.~Psaltis.
\newblock Learning approach to optical tomography.
\newblock \emph{Optica}, 2\penalty0 (6):\penalty0 517--522, 2015.

\bibitem[Li and Maiden(2018)]{Li_Maiden_2018}
Peng Li and Andrew Maiden.
\newblock Multi-slice ptychographic tomography.
\newblock \emph{Scientific Reports}, 8\penalty0 (1):\penalty0 2049, Feb 2018.
\newblock ISSN 2045-2322.
\newblock \doi{10/gfz5vw}.

\bibitem[Maiden et~al.(2012{\natexlab{a}})Maiden, Humphry, and
  Rodenburg]{Maiden_Humphry_Rodenburg_2012}
A.~M. Maiden, M.~J. Humphry, and J.~M. Rodenburg.
\newblock Ptychographic transmission microscopy in three dimensions using a
  multi-slice approach.
\newblock \emph{J. Opt. Soc. Am. A. Opt. Image Sci. Vis.}, 29\penalty0
  (8):\penalty0 1606–14, Aug 2012{\natexlab{a}}.
\newblock ISSN 1520-8532.

\bibitem[Suzuki et~al.(2014)Suzuki, Furutaku, Shimomura, Yamauchi, Kohmura,
  Ishikawa, and Takahashi]{suzuki2014high}
A.~Suzuki, S.~Furutaku, K.~Shimomura, K.~Yamauchi, Y.~Kohmura, T.~Ishikawa, and
  Y.~Takahashi.
\newblock High-resolution multislice x-ray ptychography of extended thick
  objects.
\newblock \emph{Physical review letters}, 112\penalty0 (5):\penalty0 053903,
  2014.

\bibitem[Shimomura et~al.(2015)Shimomura, Suzuki, Hirose, and
  Takahashi]{Shimomura_Suzuki_Hirose_Takahashi_2015}
K.~Shimomura, A.~Suzuki, M.~Hirose, and Y.~Takahashi.
\newblock Precession x-ray ptychography with multislice approach.
\newblock \emph{Physical Review B}, 91\penalty0 (21):\penalty0 214114, Jun
  2015.
\newblock \doi{10/gfz5vs}.

\bibitem[Tsai et~al.(2016)Tsai, Usov, Diaz, Menzel, and
  Guizar-Sicairos]{Tsai_Usov_Diaz_Menzel_Guizar-Sicairos_2016}
E.~H.~R. Tsai, I.~Usov, A.~Diaz, A.~Menzel, and M.~Guizar-Sicairos.
\newblock X-ray ptychography with extended depth of field.
\newblock \emph{Optics Express}, 24\penalty0 (25):\penalty0 29089–29108, Dec
  2016.
\newblock ISSN 1094-4087.
\newblock \doi{10/gc5qcg}.

\bibitem[\"Ozt\"urk et~al.(2018)\"Ozt\"urk, Yan, He, Ge, Dong, Lin, Nazaretski,
  Robinson, Chu, and
  Huang]{Öztürk_Yan_He_Ge_Dong_Lin_Nazaretski_Robinson_Chu_Huang_2018}
H.~\"Ozt\"urk, H.~Yan, Y.~He, M.~Ge, Z.~Dong, M.~Lin, E.~Nazaretski, I.~K.
  Robinson, Y.~S. Chu, and X.~Huang.
\newblock Multi-slice ptychography with large numerical aperture multilayer
  laue lenses.
\newblock \emph{Optica}, 5\penalty0 (5):\penalty0 601–607, May 2018.
\newblock ISSN 2334-2536.
\newblock \doi{10/gf2twh}.

\bibitem[Van~den Broek and Koch(2012)]{van2012method}
W.~Van~den Broek and C.~T Koch.
\newblock Method for retrieval of the three-dimensional object potential by
  inversion of dynamical electron scattering.
\newblock \emph{Physical review letters}, 109\penalty0 (24):\penalty0 245502,
  2012.

\bibitem[Van~den Broek and Koch(2013)]{Van_den_Broek_Koch_2013}
W.~Van~den Broek and C.~T. Koch.
\newblock General framework for quantitative three-dimensional reconstruction
  from arbitrary detection geometries in tem.
\newblock \emph{Phys. Rev. B}, 87\penalty0 (18):\penalty0 184108, May 2013.
\newblock \doi{10.1103/PhysRevB.87.184108}.

\bibitem[Gao et~al.(2017)Gao, Wang, Zhang, Martinez, Nellist, Pan, and
  Kirkland]{Gao_Wang_Zhang_Martinez_Nellist_Pan_Kirkland_2017}
Si~Gao, Peng Wang, Fucai Zhang, Gerardo~T. Martinez, Peter~D. Nellist, Xiaoqing
  Pan, and Angus~I. Kirkland.
\newblock Electron ptychographic microscopy for three-dimensional imaging.
\newblock \emph{Nature Communications}, 8, 2017.
\newblock \doi{10.1038/s41467-017-00150-1}.

\bibitem[Schloz et~al.(2020)Schloz, Pekin, Chen, Van~den Broek, Muller, and
  Koch]{Schloz_Pekin_Chen_Broek_Muller_Koch_2020}
M.~Schloz, T.~C. Pekin, Z.~Chen, W.~Van~den Broek, D.~A. Muller, and C.~T.
  Koch.
\newblock Overcoming information reduced data and experimentally uncertain
  parameters in ptychography with regularized optimization.
\newblock \emph{arXiv:2005.01530 [eess]}, May 2020.
\newblock URL \url{http://arxiv.org/abs/2005.01530}.

\bibitem[Tsai et~al.(2019)Tsai, Marone, and
  Guizar-Sicairos]{Tsai_Marone_Guizar-Sicairos_2019}
Esther H.~R. Tsai, Federica Marone, and Manuel Guizar-Sicairos.
\newblock Gridrec-ms: an algorithm for multi-slice tomography.
\newblock \emph{Optics Letters}, 44\penalty0 (9):\penalty0 2181–2184, May
  2019.
\newblock ISSN 1539-4794.
\newblock \doi{10/gfz5vr}.

\bibitem[Jacobsen(2018)]{Jacobsen_2018}
C.~Jacobsen.
\newblock Relaxation of the crowther criterion in multislice tomography.
\newblock \emph{Optics Letters}, 43\penalty0 (19):\penalty0 4811–4814, Oct
  2018.
\newblock \doi{10/gfz5vx}.

\bibitem[Xin and Muller(2009)]{Xin_Muller_2009}
H.~L. Xin and D.~A. Muller.
\newblock Aberration-corrected adf-stem depth sectioning and prospects for
  reliable 3d imaging in s/tem.
\newblock \emph{Journal of Electron Microscopy}, 58\penalty0 (3):\penalty0
  157–165, Jun 2009.
\newblock ISSN 0022-0744.
\newblock \doi{10/fndbg5}.

\bibitem[Gilles et~al.(2018)Gilles, Nashed, Du, Jacobsen, and
  Wild]{Gilles_Nashed_Du_Jacobsen_Wild_2018}
M.~A. Gilles, Y.~S.~G. Nashed, M.~Du, C.~Jacobsen, and S.~M. Wild.
\newblock 3d x-ray imaging of continuous objects beyond the depth of focus
  limit.
\newblock \emph{Optica}, 5\penalty0 (9):\penalty0 1078–1086, Sep 2018.
\newblock ISSN 2334-2536.
\newblock \doi{10/gfmhzc}.

\bibitem[Du et~al.(2020)Du, Nashed, Kandel, Gürsoy, and
  Jacobsen]{Du_Nashed_Kandel_Gürsoy_Jacobsen_2020}
M.~Du, Y.~S.~G. Nashed, S.~Kandel, D.~Gürsoy, and C.~Jacobsen.
\newblock Three dimensions, two microscopes, one code: Automatic
  differentiation for x-ray nanotomography beyond the depth of focus limit.
\newblock \emph{Science Advances}, 6\penalty0 (13):\penalty0 eaay3700, Mar
  2020.
\newblock \doi{10.1126/sciadv.aay3700}.

\bibitem[Ren et~al.(2020)Ren, Ophus, Chen, and Waller]{ren2020multiple}
D.~Ren, C.~Ophus, M.~Chen, and L.~Waller.
\newblock A multiple scattering algorithm for three dimensional phase contrast
  atomic electron tomography.
\newblock \emph{Ultramicroscopy}, 208:\penalty0 112860, 2020.

\bibitem[Fujimoto(1959)]{fujimoto1959dynamical}
F.~Fujimoto.
\newblock Dynamical theory of electron diffraction in {L}aue-case, {I}. general
  theory.
\newblock \emph{Journal of the Physical Society of Japan}, 14\penalty0
  (11):\penalty0 1558--1568, 1959.

\bibitem[Sturkey(1962)]{sturkey1962calculation}
L.~Sturkey.
\newblock The calculation of electron diffraction intensities.
\newblock \emph{Proceedings of the Physical Society}, 80\penalty0 (2):\penalty0
  321, 1962.

\bibitem[Ophus(2017)]{ophus2017fast}
C.~Ophus.
\newblock A fast image simulation algorithm for scanning transmission electron
  microscopy.
\newblock \emph{Advanced structural and chemical imaging}, 3\penalty0
  (1):\penalty0 13, 2017.

\bibitem[Brown et~al.(2019)Brown, Ciston, and Ophus]{brown2019linear}
H.~G. Brown, J.~Ciston, and C.~Ophus.
\newblock Linear-scaling algorithm for rapid computation of inelastic
  transitions in the presence of multiple electron scattering.
\newblock \emph{Physical Review Research}, 1\penalty0 (3):\penalty0 033186,
  2019.

\bibitem[Spence(1998)]{spence1998direct}
J.C.H. Spence.
\newblock Direct inversion of dynamical electron diffraction patterns to
  structure factors.
\newblock \emph{Acta Crystallographica Section A: Foundations of
  Crystallography}, 54\penalty0 (1):\penalty0 7--18, 1998.

\bibitem[Allen et~al.(1998)Allen, Josefsson, and
  Leeb]{Allen_Josefsson_Leeb_1998}
L.~J. Allen, T.~W. Josefsson, and H.~Leeb.
\newblock Obtaining the crystal potential by inversion from electron scattering
  intensities.
\newblock \emph{Acta Crystallographica Section A: Foundations of
  Crystallography}, 54\penalty0 (44):\penalty0 388–398, Jul 1998.
\newblock ISSN 0108-7673.
\newblock \doi{10/bg6bgk}.

\bibitem[Allen et~al.(2000)Allen, Faulkner, and Leeb]{allen2000inversion}
L.~J. Allen, H.~M.~L. Faulkner, and H.~Leeb.
\newblock Inversion of dynamical electron diffraction data including
  absorption.
\newblock \emph{Acta Crystallographica Section A: Foundations of
  Crystallography}, 56\penalty0 (2):\penalty0 119--126, 2000.

\bibitem[Donatelli and Spence(2020)]{Donatelli_Spence_2020}
J.~J. Donatelli and J.~C.~H. Spence.
\newblock Inversion of many-beam bragg intensities for phasing by iterated
  projections: Removal of multiple scattering artifacts from diffraction data.
\newblock \emph{Physical Review Letters}, 125\penalty0 (6):\penalty0 065502,
  Aug 2020.
\newblock \doi{10.1103/PhysRevLett.125.065502}.

\bibitem[Rez(1999)]{Rez_1999}
P.~Rez.
\newblock Schemes to determine the crystal potential under dynamical conditions
  using voltage variation.
\newblock \emph{Acta Crystallographica Section A: Foundations of
  Crystallography}, 55\penalty0 (2):\penalty0 160--167, 1999.

\bibitem[Wang et~al.(2016)Wang, Pennington, and
  Koch]{Wang_Pennington_Koch_2016}
F.~Wang, R.~S. Pennington, and C.~T. Koch.
\newblock \emph{Physical Review Letters}, \penalty0 (1):\penalty0 015501, Jun
  2016.

\bibitem[Findlay(2005)]{findlay2005quantitative}
S.~D. Findlay.
\newblock Quantitative structure retrieval using scanning transmission electron
  microscopy.
\newblock \emph{Acta Crystallographica Section A: Foundations of
  Crystallography}, 61\penalty0 (4):\penalty0 397--404, 2005.

\bibitem[Brown et~al.(2018)Brown, Chen, Weyland, Ophus, Ciston, Allen, and
  Findlay]{brown2018structure}
H.~G. Brown, Z.~Chen, M.~Weyland, C.~Ophus, J.~Ciston, L.~J. Allen, and S.~D.
  Findlay.
\newblock Structure retrieval at atomic resolution in the presence of multiple
  scattering of the electron probe.
\newblock \emph{Physical Review Letters}, 121\penalty0 (26):\penalty0 266102,
  2018.

\bibitem[Popoff et~al.(2010)Popoff, Lerosey, Carminati, Fink, Boccara, and
  Gigan]{Popoff_Lerosey_Carminati_Fink_Boccara_Gigan_2010}
S.~M. Popoff, G.~Lerosey, R.~Carminati, M.~Fink, A.~C. Boccara, and S.~Gigan.
\newblock Measuring the transmission matrix in optics: An approach to the study
  and control of light propagation in disordered media.
\newblock \emph{Physical Review Letters}, 104\penalty0 (10):\penalty0 100601,
  Mar 2010.
\newblock ISSN 0031-9007, 1079-7114.
\newblock \doi{10/b6hg8x}.

\bibitem[Kim et~al.(2012)Kim, Choi, Yoon, Choi, Kim, Park, and
  Choi]{Kim_Choi_Yoon_Choi_Kim_Park_Choi_2012}
Moonseok Kim, Youngwoon Choi, Changhyeong Yoon, Wonjun Choi, Jaisoon Kim,
  Q.-Han Park, and Wonshik Choi.
\newblock Maximal energy transport through disordered media with the
  implementation of transmission eigenchannels.
\newblock \emph{Nature Photonics}, 6\penalty0 (9):\penalty0 581–585, Sep
  2012.
\newblock ISSN 1749-4893.
\newblock \doi{10/gbbsw9}.

\bibitem[Metzler et~al.(2017)Metzler, Sharma, Nagesh, Baraniuk, Cossairt, and
  Veeraraghavan]{Metzler_Sharma_Nagesh_Baraniuk_Cossairt_Veeraraghavan_2017}
C.~A. Metzler, M.~K. Sharma, S.~Nagesh, R.~G. Baraniuk, O.~Cossairt, and
  A.~Veeraraghavan.
\newblock Coherent inverse scattering via transmission matrices: Efficient
  phase retrieval algorithms and a public dataset.
\newblock In \emph{2017 IEEE International Conference on Computational
  Photography (ICCP)}, page 1–16, May 2017.
\newblock \doi{10/ggh89w}.

\bibitem[Dr\'emeau et~al.(2015)Dr\'emeau, Liutkus, Martina, Katz, Sch\"ulke,
  Krzakala, Gigan, and
  Daudet]{Dremeau_Liutkus_Martina_Katz_Schulke_Krzakala_Gigan_Daudet_2015}
A.~Dr\'emeau, A.~Liutkus, D.~Martina, O.~Katz, C.~Sch\"ulke, F.~Krzakala,
  S.~Gigan, and L.~Daudet.
\newblock Reference-less measurement of the transmission matrix of a highly
  scattering material using a {DMD} and phase retrieval techniques.
\newblock \emph{Optics Express}, 23\penalty0 (9):\penalty0 11898–11911, May
  2015.
\newblock ISSN 1094-4087.
\newblock \doi{10/ggm6g2}.

\bibitem[Rajaei et~al.(2016)Rajaei, Tramel, Gigan, Krzakala, and
  Daudet]{Rajaei_Tramel_Gigan_Krzakala_Daudet_2016}
B.~Rajaei, E.~W. Tramel, S.~Gigan, F.~Krzakala, and L.~Daudet.
\newblock Intensity-only optical compressive imaging using a multiply
  scattering material and a double phase retrieval approach.
\newblock In \emph{2016 IEEE International Conference on Acoustics, Speech and
  Signal Processing (ICASSP)}, page 4054–4058, Mar 2016.
\newblock \doi{10/ggm6g3}.

\bibitem[Yu et~al.(2013)Yu, Hillman, Choi, Lee, Feld, Dasari, and
  Park]{Yu_Hillman_Choi_Lee_Feld_Dasari_Park_2013}
H.~Yu, T.~R. Hillman, W.~Choi, J.~O. Lee, M.~S. Feld, R.~R. Dasari, and
  Y.~Park.
\newblock Measuring large optical transmission matrices of disordered media.
\newblock \emph{Physical Review Letters}, 111\penalty0 (15):\penalty0 153902,
  Oct 2013.
\newblock \doi{10/f5mq6f}.

\bibitem[Kirkland(2020)]{kirkland2020advanced}
E.~J. Kirkland.
\newblock \emph{Advanced Computing in Electron Microscopy, 3rd Edition}.
\newblock Springer, 2020.

\bibitem[Rotter and Gigan(2017)]{Rotter_Gigan_2017}
S.~Rotter and S.~Gigan.
\newblock Light fields in complex media: Mesoscopic scattering meets wave
  control.
\newblock \emph{Reviews of Modern Physics}, 89\penalty0 (1):\penalty0 015005,
  Mar 2017.
\newblock \doi{10/f9t5fk}.

\bibitem[Miao et~al.(1998)Miao, Sayre, and Chapman]{Miao_Sayre_Chapman_1998}
J.~Miao, D.~Sayre, and H.~N. Chapman.
\newblock Phase retrieval from the magnitude of the fourier transforms of
  nonperiodic objects.
\newblock \emph{JOSA A}, 15\penalty0 (6):\penalty0 1662–1669, Jun 1998.
\newblock ISSN 1520-8532.
\newblock \doi{10/fw83fw}.

\bibitem[Fienup(1982)]{Fienup_1982}
J~R Fienup.
\newblock Phase retrieval algorithms: a comparison.
\newblock \emph{Appl. Opt.}, 21\penalty0 (15):\penalty0 2758–69, Aug 1982.
\newblock ISSN 0003-6935.

\bibitem[Shechtman et~al.(2015)Shechtman, Eldar, Cohen, Chapman, Miao, and
  Segev]{Shechtman_Eldar_Cohen_Chapman_Miao_Segev_2015}
Y.~Shechtman, Y.C. Eldar, O.~Cohen, H.N. Chapman, Jianwei Miao, and M.~Segev.
\newblock Phase retrieval with application to optical imaging: A contemporary
  overview.
\newblock \emph{IEEE Signal Processing Magazine}, 32\penalty0 (3):\penalty0
  87–109, May 2015.
\newblock ISSN 1053-5888.
\newblock \doi{10.1109/MSP.2014.2352673}.

\bibitem[Candes et~al.(2013)Candes, Strohmer, and Voroninski]{PhaseLift_2013}
E.~J. Candes, T.~Strohmer, and V.~Voroninski.
\newblock Phaselift: Exact and stable signal recovery from magnitude
  measurements via convex programming.
\newblock \emph{Communications on Pure and Applied Mathematics}, 66\penalty0
  (8):\penalty0 1241--1274, 2013.

\bibitem[Waldspurger et~al.(2015)Waldspurger, d’Aspremont, and
  Mallat]{Waldspurger_d’Aspremont_Mallat_2015}
I.~Waldspurger, A.~d’Aspremont, and S.~Mallat.
\newblock Phase recovery, maxcut and complex semidefinite programming.
\newblock \emph{Mathematical Programming}, 149\penalty0 (1):\penalty0 47–81,
  Feb 2015.
\newblock ISSN 1436-4646.
\newblock \doi{10.1007/s10107-013-0738-9}.

\bibitem[{Bostan} et~al.(2018){Bostan}, {Soltanolkotabi}, {Ren}, and
  {Waller}]{Bostan_Soltanolkotabi_Ren_Waller_2018}
E.~{Bostan}, M.~{Soltanolkotabi}, D.~{Ren}, and L.~{Waller}.
\newblock Accelerated wirtinger flow for multiplexed fourier ptychographic
  microscopy.
\newblock In \emph{2018 25th IEEE International Conference on Image Processing
  (ICIP)}, pages 3823--3827, 2018.

\bibitem[Nikitin et~al.(2019)Nikitin, Aslan, Yao, Bi{\c{c}}er, Leyffer, Mokso,
  and G{\"u}rsoy]{nikitin2019photon}
V.~Nikitin, S.~Aslan, Y.~Yao, T.~Bi{\c{c}}er, S.~Leyffer, R.~Mokso, and
  D.~G{\"u}rsoy.
\newblock Photon-limited ptychography of 3d objects via bayesian
  reconstruction.
\newblock \emph{OSA Continuum}, 2\penalty0 (10):\penalty0 2948--2968, 2019.

\bibitem[Yeh et~al.(2015)Yeh, Dong, Zhong, Tian, Chen, Tang, Soltanolkotabi,
  and Waller]{Yeh_Dong_Zhong_Tian_Chen_Tang_Soltanolkotabi_Waller_2015}
L.-H. Yeh, J.~Dong, J.~Zhong, L.~Tian, M.~Chen, G.~Tang, M.~Soltanolkotabi, and
  L.~Waller.
\newblock Experimental robustness of fourier ptychography phase retrieval
  algorithms.
\newblock \emph{Optics Express}, 23\penalty0 (26):\penalty0 33214, Dec 2015.
\newblock ISSN 1094-4087.
\newblock \doi{10/gc5pwb}.

\bibitem[Fannjiang and Strohmer(2020)]{Fannjiang_Strohmer_2020}
A.~Fannjiang and T.~Strohmer.
\newblock The numerics of phase retrieval.
\newblock \emph{arXiv:2004.05788 [cs, eess, math]}, Apr 2020.
\newblock URL \url{http://arxiv.org/abs/2004.05788}.

\bibitem[Parikh and Boyd(2014)]{Parikh_Boyd_2014}
N.~Parikh and S.~Boyd.
\newblock Proximal algorithms.
\newblock \emph{Found. Trends Optim.}, 1\penalty0 (3):\penalty0 127–239, Jan
  2014.
\newblock ISSN 2167-3888.
\newblock \doi{10.1561/2400000003}.

\bibitem[Wen et~al.(2012)Wen, Yang, Liu, and
  Marchesini]{Wen_Yang_Liu_Marchesini_2012}
Z.~Wen, C.~Yang, X.~Liu, and S.~Marchesini.
\newblock Alternating direction methods for classical and ptychographic phase
  retrieval.
\newblock \emph{Inverse Problems}, 28\penalty0 (11):\penalty0 115010, Oct 2012.
\newblock ISSN 0266-5611.
\newblock \doi{10/gf3q42}.

\bibitem[Verbeeck et~al.(2018)Verbeeck, B\'ech\'e, M\"uller-Caspary, Guzzinati,
  Luong, and
  Den~Hertog]{Verbeeck_Beche_Muller_Caspary_Guzzinati_Luong_Den_Hertog_2018}
J.~Verbeeck, A.~B\'ech\'e, K.~M\"uller-Caspary, G.~Guzzinati, M.~A. Luong, and
  M.~Den~Hertog.
\newblock Demonstration of a 2 × 2 programmable phase plate for electrons.
\newblock \emph{Ultramicroscopy}, 190:\penalty0 58–65, Jul 2018.
\newblock ISSN 0304-3991.
\newblock \doi{10/gdpbjd}.

\bibitem[Kong et~al.(2011)Kong, Silverman, Liu, Chitnis, Lee, and
  Chen]{Kong_Silverman_Liu_Chitnis_Lee_Chen_2011}
F.~Kong, R.~H. Silverman, L.~Liu, P.~V. Chitnis, K.~K. Lee, and Y.~C. Chen.
\newblock Photoacoustic-guided convergence of light through optically diffusive
  media.
\newblock \emph{Optics Letters}, 36\penalty0 (11):\penalty0 2053–2055, Jun
  2011.
\newblock ISSN 1539-4794.
\newblock \doi{10/dhr9j8}.

\bibitem[Maiden et~al.(2012{\natexlab{b}})Maiden, Humphry, Sarahan, Kraus, and
  Rodenburg]{Maiden_Humphry_Sarahan_Kraus_Rodenburg_2012}
A.~M. Maiden, M.~J. Humphry, M.~C. Sarahan, B.~Kraus, and J.~M. Rodenburg.
\newblock An annealing algorithm to correct positioning errors in ptychography.
\newblock \emph{Ultramicroscopy}, 120:\penalty0 64–72, Sep
  2012{\natexlab{b}}.
\newblock ISSN 0304-3991.
\newblock \doi{10.1016/j.ultramic.2012.06.001}.

\bibitem[Odstrčil et~al.(2018)Odstrčil, Menzel, and
  Guizar-Sicairos]{Odstril_Menzel_Guizar_Sicairos_2018}
M.~Odstrčil, A.~Menzel, and M.~Guizar-Sicairos.
\newblock Iterative least-squares solver for generalized maximum-likelihood
  ptychography.
\newblock \emph{Optics Express}, 26\penalty0 (3):\penalty0 3108–3123, Feb
  2018.
\newblock ISSN 1094-4087.
\newblock \doi{10/gcx53m}.

\bibitem[Rana et~al.(2020)Rana, Zhang, Pham, Yuan, Lo, Jiang, Osher, and
  Miao]{Rana_Zhang_Pham_Yuan_Lo_Jiang_Osher_Miao_2020}
A.~Rana, J.~Zhang, M.~Pham, A.~Yuan, Yuan-Hung Lo, H.~Jiang, S.~J. Osher, and
  J.~Miao.
\newblock Potential of attosecond coherent diffractive imaging.
\newblock \emph{Physical Review Letters}, 125\penalty0 (8):\penalty0 086101,
  Aug 2020.
\newblock \doi{10.1103/PhysRevLett.125.086101}.

\bibitem[Thibault and Menzel(2013)]{Thibault_Menzel_2013}
P.~Thibault and A.~Menzel.
\newblock Reconstructing state mixtures from diffraction measurements.
\newblock \emph{Nature}, 494\penalty0 (7435):\penalty0 68–71, Feb 2013.
\newblock ISSN 1476-4687.
\newblock \doi{10.1038/nature11806}.

\bibitem[Chen et~al.(2020)Chen, Odstrcil, Jiang, Han, Chiu, Li, and
  Muller]{Chen_Odstrcil_Jiang_Han_Chiu_Li_Muller_2020}
Z.~Chen, M.~Odstrcil, Y.~Jiang, Y.~Han, M.-H. Chiu, L.-J. Li, and D.~A. Muller.
\newblock Mixed-state electron ptychography enables sub-angstrom resolution
  imaging with picometer precision at low dose.
\newblock \emph{Nature Communications}, 11\penalty0 (11):\penalty0 2994, Jun
  2020.
\newblock ISSN 2041-1723.
\newblock \doi{10/gg2fbw}.

\end{thebibliography}
\bibliographystyle{unsrtnat}



\newpage
\section{Derivation of the adjoint S-matrix measurement operator}
\begin{figure*}[ht]
    \includegraphics[width=\textwidth]{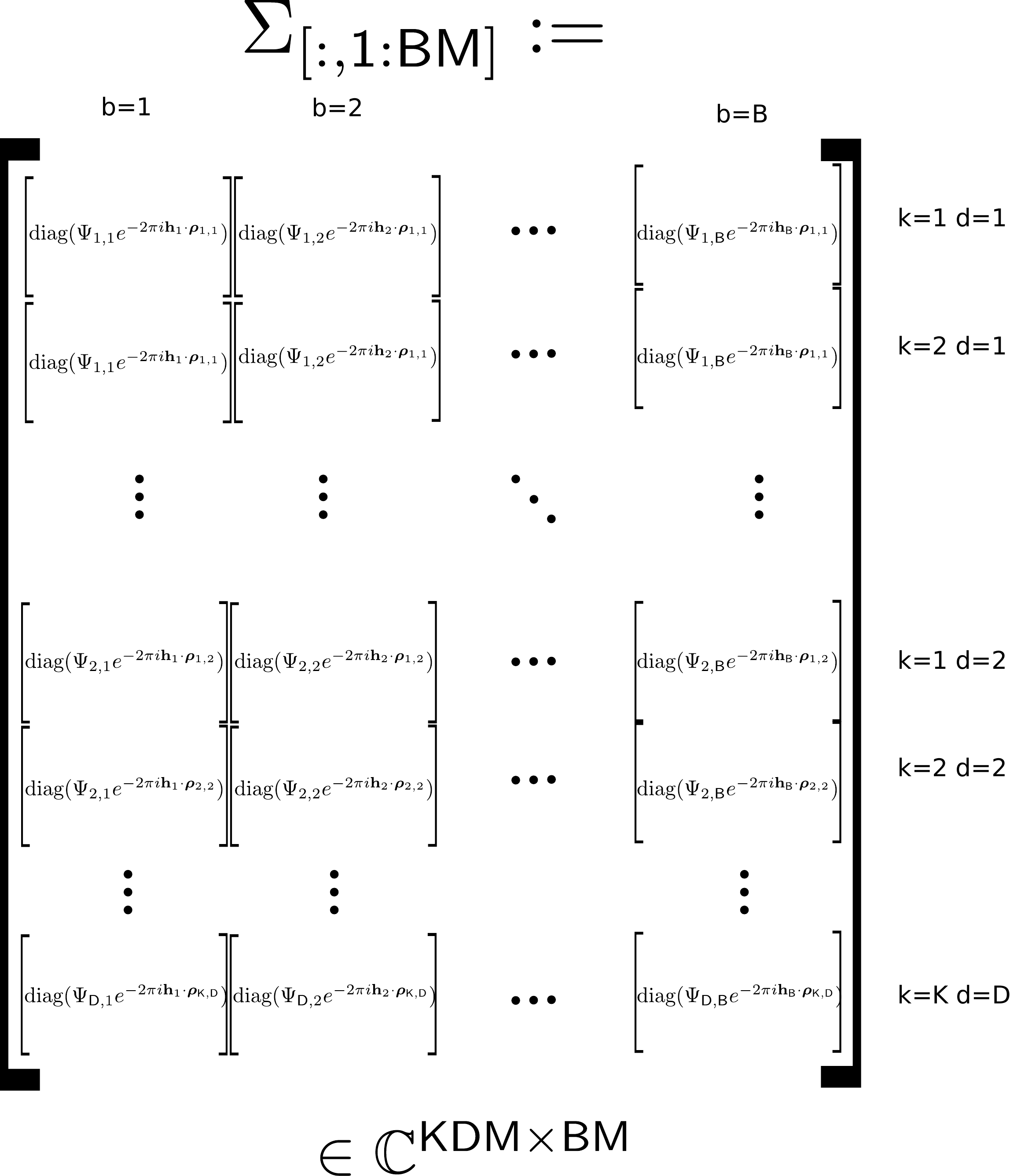}
    \caption{The first $\mathsf{BM}$ columns of $\mathbf{\Sigma}$ written out explicitly with diagonal matrix blocks of size $\mathsf{M}\times\mathsf{M}$.}
    \label{fig:sigma_operator}
\end{figure*}

We derive the adjoint operators $\mathcal{A}_{\mathsf{k},\mathsf{d}}^{\mathcal{S}_{\mathsf{b}}\,\dagger}(\mathbf{z})$ and $\mathcal{A}_{\mathsf{k},\mathsf{d}}^{\Psi_{\textsf{d},\textsf{b}}\,\dagger}(\mathbf{z})$ with matrix algebra.
In matrix notation, the \smatrix{} forward model to generate $\mathbf{I}\in \mathbb{R}^{\mathsf{KDM}}$ with $\mathsf{M} = \mathsf{M}_1\cdot\mathsf{M}_2$ from $\bm{\mathcal{S}} \in \mathbb{C}^{\mathsf{BN}}$ with $\mathsf{N} = \mathsf{N}_1\cdot\mathsf{N}_2$ can be written as $\bm{\mathcal{A}}_{\mathcal{S}} : \mathbb{C}^{\mathsf{BN}} \rightarrow \mathbb{C}^{\mathsf{KDM}}$
\begin{equation}
    \mathbf{I} = \left|\mathbf{F}\mathbf{\Sigma}\mathbf{C}\bm{\mathcal{S}}\right|^2 = \left|\bm{\mathcal{A}}_{\mathcal{S}}\bm{\mathcal{S}}\right|^2,
\end{equation}
with $\mathbf{F} \in \mathbb{C}^{\mathsf{KDM} \times \mathsf{KDM}}$ a block-diagonal matrix representing a batched Fourier transform acting on $\mathsf{KD}$ exit waves, $\mathbf{C} \in \mathbb{R}^{\mathsf{KDBM} \times \mathsf{BN}}$ the cropping matrix that extracts $\mathsf{KD}$ patches centered at the scanning positions out of the $\mathsf{B}$ beams of the \smatrix{}, and $\mathbf{\Sigma} \in \mathbb{C}^{\mathsf{KDM} \times \mathsf{KDBM}}$ the coherent summation operator over all beams. Written out in block matrices with diagonal entries, $\mathbf{\Sigma}$ is shown in Fig. \ref{fig:sigma_operator}.
The adjoint (hermitian transpose) operator $\bm{\mathcal{A}}^{\dagger}_{\mathcal{S}}$ is then
\begin{equation}
    \bm{\mathcal{A}}^{\dagger}_{\mathcal{S}} = \mathbf{C}^{T}\mathbf{\Sigma}^{\dagger}\mathbf{F}^{\dagger},
\end{equation}
which, written out for a single diffraction pattern with defocus index $\mathsf{d}$ and position index $\mathsf{k}$ is
\begin{equation}
    \mathcal{A}_{\mathsf{k},\mathsf{d}}^{\mathcal{S}_{\mathsf{b}}\,\dagger}(\mathbf{z}) = \mathbf{C}_{\mathsf{k},\mathsf{d}}^{T}\left[\Psi_{\textsf{d},\textsf{b}}^*e^{2 \pi i\vec{h}_{\mathsf{b}}\cdot\bm{\rho}_{\mathsf{k},\mathsf{d}}}\finv[\vec{q}]{\mathbf{z_{\mathsf{k},\mathsf{d}}}}\right] 
    \label{equ:adjoint_S}
\end{equation}
In the same vein, the forward model to generate ${\mathbf{I}\in \mathbb{R}^{\mathsf{KDM}}}$ from $\bm{\Psi} \in \mathbb{C}^{\mathsf{DB}}$ can be written as ${\bm{\mathcal{A}}_{\mathbf{\Psi}} : \mathbb{C}^{\mathsf{DB}} \rightarrow \mathbb{C}^{\mathsf{KDM}}}$
\begin{equation}
    \mathbf{I} = \left|\mathbf{F}\mathbf{\Sigma}_{\mathbf{\Psi}}\mathbf{\Psi}\right|^2 = \left|\bm{\mathcal{A}}_{\mathbf{\Psi}}\mathbf{\Psi}\right|^2
\end{equation}
The adjoint operator $\bm{\mathcal{A}}^{\dagger}_{\mathbf{\Psi}}$ is then
\begin{equation}
    \bm{\mathcal{A}}^{\dagger}_{\mathbf{\Psi}} = \mathbf{\Sigma}^{\dagger}_{\mathbf{\Psi}}\mathbf{F}^{\dagger},
\end{equation}
which, written out for a single diffraction pattern with defocus index $\mathsf{d}$ and position index $\mathsf{k}$ is
\begin{eqnarray}
    \mathcal{A}_{\mathsf{k},\mathsf{d}}^{\Psi_{\textsf{d},\textsf{b}}\,\dagger}(\mathbf{z}) =&& \frac{1}{\mathsf{M_1}\mathsf{M_2}} \sum_{\mathsf{m_1}}^{\mathsf{M_1}}\sum_{\mathsf{m_2}}^{\mathsf{M_2}}\nonumber\\&& \left[\sum_{\mathsf{k=1}}^{\textsf{K}} \left[\mathbf{C}_{\mathsf{k},\mathsf{d}}\mathcal{S}\right]^*_{\mathsf{b}}e^{2 \pi i\vec{h}_{\mathsf{b}}\cdot\bm{\rho}_{\mathsf{k},\mathsf{d}}}\finv[\vec{q}]{\mathbf{z}_{\mathsf{k},\mathsf{d}}}\right]_{\mathsf{m_1},\mathsf{m_2}} 
    \label{equ:adjoint_Psi}
\end{eqnarray}

\end{document}